\pdfoutput=1

\documentclass[11pt,twoside,a4paper,cmspaper,final,collab]{cms-tdr}
\def\svnVersion{300973}\def\cmsCernNo{2013-037}\def\cmsCernDate{\today}\def\cmsMessage{Published in Physics Letters B as \href{http://dx.doi.org/10.1016/j.physletb.2015.07.053}{\doi{10.1016/j.physletb.2015.07.053.}}}

\begin{document}\cmsNoteHeader{HIG-14-005}

\hyphenation{had-ron-i-za-tion}
\hyphenation{cal-or-i-me-ter}
\hyphenation{de-vices}

\RCS$Revision: 300870 $
\RCS$HeadURL: svn+ssh://svn.cern.ch/reps/tdr2/papers/HIG-14-005/trunk/HIG-14-005.tex $
\RCS$Id: HIG-14-005.tex 300870 2015-08-20 16:26:42Z cjessop $
\newlength\cmsFigWidth
\providecommand{\tauh}{\ensuremath{\Pgt_\mathrm{h}}\xspace}
\providecommand{\NA}{---}
\providecommand{\MT}{\ensuremath{M_\mathrm{T}}\xspace}
\ifthenelse{\boolean{cms@external}}{\newcommand{\cmsTable}[2]{\resizebox{#1}{!}{#2}}}{\newcommand{\cmsTable}[2]{\relax#2}}

\cmsNoteHeader{HIG-14-005 v1.0}
\title{Search for lepton-flavour-violating decays of the Higgs boson}

\date{\today}

\abstract{The first direct search for lepton-flavour-violating decays of the recently discovered Higgs boson (H) is described. The search is performed in the $\PH \to \Pgm \Pgt_{\Pe}$ and $\PH \to \Pgm \tauh$  channels, where  $\Pgt_{\Pe}$ and   $\tauh$ are tau leptons reconstructed in the electronic  and  hadronic decay channels, respectively.
The data sample used in this search was collected in pp collisions at a centre-of-mass
energy of $\sqrt{s}=8$\TeV   with the CMS experiment at the CERN LHC and corresponds to an
integrated luminosity of 19.7\fbinv.
The sensitivity of the search is an order of magnitude better than the existing indirect limits. A slight excess of signal events with
a significance of 2.4 standard deviations is observed. The  $p$-value of this excess at $M_{\PH}=125$\GeV is 0.010.
The best fit branching fraction is $\mathcal{B}(\PH \to \Pgm \Pgt )=(0.84^{+0.39}_{-0.37})\%$. A constraint on the branching fraction,
$\mathcal{B}(\PH \to \Pgm \Pgt )<1.51\%$ at 95\% confidence level is set.
This limit is subsequently used to constrain the $\mu$-$\tau$  Yukawa couplings to be less than
$3.6\times 10^{-3}$.
}

\hypersetup{%
pdfauthor={CMS Collaboration},%
pdftitle={Search for lepton-flavour-violating decays of the Higgs boson},%
pdfsubject={CMS},%
pdfkeywords={CMS, physics, Higgs, muons, taus, lepton-flavour-violation}}

\maketitle
\section{Introduction}
The discovery of the Higgs boson ($\PH$)~\cite{Aad:2012tfa,Chatrchyan:2012ufa,Chatrchyan:2013lba} has generated great interest in
exploring  its properties. In the standard model (SM), lepton-flavour-violating (LFV) decays are
forbidden if the theory is to be renormalizable~\cite{Harnik:2012pb}. If this requirement is relaxed, so the
theory is valid only to a finite mass scale, then LFV couplings may be introduced. LFV decays can
 occur naturally in models with more than one Higgs doublet without abandoning
renormalizability~\cite{PhysRevLett.38.622}. They also arise in supersymmetric models~\cite{DiazCruz:1999xe,Han:2000jz,Arhrib:2012ax,Arana-Catania:2013xma}, composite Higgs boson
models~\cite{Agashe:2009di,Azatov:2009na}, models with flavour symmetries~\cite{Ishimori:2010au},
Randall--Sundrum models~\cite{Casagrande:2008hr,Buras:2009ka,Perez:2008ee}, and many others~\cite{Blanke:2008zb,Giudice:2008uua,AguilarSaavedra:2009mx,Albrecht:2009xr,Goudelis:2011un,McKeen:2012av,Arganda:2004bz,Arganda:2014dta}. The presence of LFV couplings
would allow $\Pgm \to \Pe$, $\Pgt \to \Pgm$ and $\Pgt \to \Pe$ transitions to proceed via a virtual Higgs boson~\cite{McWilliams:1980kj,Shanker:1981mj}. The experimental limits  on these have recently been translated into constraints on the branching fractions
$\mathcal{B}(\PH \to \Pe \Pgm,\Pgm \Pgt, \Pe \Pgt)$~\cite{Blankenburg:2012ex,Harnik:2012pb}.
The $\Pgm \to \Pe$ transition is strongly constrained by
null search results for $\Pgm \to \Pe \gamma$~\cite{Agashe:2014kda},
$\mathcal{B}(\PH \to \Pgm \Pe) < \mathcal{O}(10^{-8})$. However, the constraints on $\Pgt \to \Pgm$ and $\Pgt \to \Pe $
are much less stringent. These come from  searches for $\Pgt \to \Pgm \gamma$~\cite{Kanemura:2005hr,Davidson:2010xv}
and other rare $\Pgt$ decays~\cite{Celis:2013xja}, $\Pgt \to \Pe \gamma$, $\Pgm$ and $\Pe$ $g-2$ measurements~\cite{Agashe:2014kda}. Exclusion limits
on the electron and muon electric dipole moments~\cite{Barr:1990vd} also provide complementary constraints. These lead to the much less restrictive limits: $\mathcal{B}(\PH \to \Pgm \Pgt) < \mathcal{O}(10\%)$, $\mathcal{B}(\PH \to \Pe \Pgt) < \mathcal{O}(10\%)$. The observation of the Higgs boson offers the possibility of sensitive direct searches for
LFV Higgs boson  decays. To date no dedicated searches have been performed. However,
a theoretical reinterpretation of the  ATLAS $\PH \to \Pgt \Pgt$ search
results in terms of LFV decays by an independent group has been used to set limits at the  95\% confidence level (CL) of $\mathcal{B}(\PH \to \Pgm \Pgt) <  13\%$,
$\mathcal{B}(\PH \to \Pe \Pgt) < 13\%$~\cite{Harnik:2012pb}.

This letter describes a search  for a LFV decay of a Higgs boson with $\MH=125$\GeV  at the CMS experiment. The  2012 dataset collected at a centre-of-mass energy  of $\sqrt{s}=8$\TeV corresponding to an integrated
luminosity of 19.7 $\fbinv$ is used. The search is performed in  two  channels,  $\PH \to \Pgm \Pgt_{\Pe}$ and
$\PH \to \Pgm \tauh$, where $\Pgt_{\Pe}$ and $\tauh$ are tau leptons  reconstructed in the electronic and
hadronic decay channels, respectively.  The  signature is very similar to
the SM $\PH \to \Pgt_{\Pgm}\Pgt_{\Pe}$ and $\PH \to \Pgt_{\Pgm}\tauh$ decays, where $\Pgt_{\Pgm}$ is a tau lepton decaying muonically, which have been studied by CMS in Refs.~\cite{Chatrchyan:2014vua,Chatrchyan:2014nva} and ATLAS in Ref.~\cite{Aad:2015vsa},
but with some significant kinematic differences. The $\Pgm$ comes promptly from the LFV \PH decay and tends to have a larger momentum
than in the SM case. There is only one tau lepton so there are typically fewer neutrinos in the decay. They are  highly Lorentz boosted and tend
to be collinear with the visible $\Pgt$ decay products.

The two channels are divided into
categories based on the number of jets in order to separate the different \PH boson production mechanisms.
The signal sensitivity is enhanced by using different selection criteria for each category.
The dominant production
mechanism is gluon-gluon fusion but there is also a significant contribution from vector boson fusion which is enhanced
by requiring jets to be present in the event.
The dominant background in the   $\PH \to \Pgm \Pgt_{\Pe}$ channel is
$ \cPZ \to \Pgt \Pgt$. Other much smaller backgrounds come from  misidentified  leptons  in \PW+jets, QCD multijets
and $\ttbar$ events. In the  $\PH \to \Pgm \tauh$ channel the dominant background arises from
misidentified $\Pgt$ leptons in \PW+jets, QCD multijets and $\ttbar$ events. Less
significant backgrounds  come from $\cPZ \to \Pgt\Pgt$ and \cPZ+jets. The principal backgrounds are estimated using data.
There is also a small background from  SM \PH decays which is estimated with simulation.
The presence or absence of a signal is established by fitting a mass distribution for signal and background using the asymptotic CL$_{\text{s}}$ criterion~\cite{Junk,Read2}.
A ``blind'' analysis was performed. The data in the  signal region were not studied  until the selection criteria had been fixed and the background estimate finalized.

\section{Detector and data sets}

A detailed description of the CMS detector, together with a description of the coordinate system used and the relevant kinematic variables, can be found
in ref.~\cite{CMS-JINST}. The momenta of charged particles are
measured with a silicon pixel and strip tracker that covers the
pseudorapidity range $\abs{\eta} < 2.5$ and is inside a 3.8\unit{T}
axial magnetic field. Surrounding the tracker are a
lead tungstate crystal electromagnetic calorimeter
(ECAL) and a brass/scintillator hadron calorimeter,
both consisting of a barrel assembly and two endcaps that extend to a pseudorapidity range of $\abs{\eta} < 3.0$. A
steel/quartz-fiber Cherenkov forward detector extends the calorimetric
coverage to $\abs{\eta} < 5.0$. The outermost component of the CMS detector is the
muon system, consisting of gas-ionization detectors placed in the
steel flux-return yoke of the magnet
to measure the momenta of muons traversing  the detector. The two-level CMS trigger system selects events of interest for
permanent storage. The first trigger level,
composed of custom hardware processors, uses information from the
calorimeters and muon detectors to select events in less than 3.2\mus.
The high-level trigger software algorithms, executed on a farm of
commercial processors, further reduce the
event rate using information from all detector subsystems.

The $\PH \to \Pgm \tauh$ channel selection begins by  requiring a single $\Pgm$
trigger  with a transverse momentum threshold  $\pt^{\Pgm}>24$\GeV in the pseudorapidity range $\abs{\eta} < 2.1$,
while the $\PH \to \Pgm \Pgt_{\Pe}$ channel requires a $\Pgm$-$\Pe$ trigger with $\pt$ thresholds
of 17\GeV ($\abs{\eta} < 2.4$) for the $\Pgm$  and 8\GeV ($\abs{\eta} < 2.5$) for the $\Pe$. Loose $\Pe$ and $\Pgm$
identification criteria are applied at the trigger level. The leptons are
also required
to be isolated from other tracks and calorimeter energy deposits to maintain an acceptable trigger
rate.

Simulated samples of signal and background events are produced using various Monte Carlo (MC) event generators, with the CMS detector response modeled with \GEANTfour~\cite{GEANT4}.
Higgs bosons are produced in proton-proton collisions predominantly by gluon-gluon fusion, but also by vector boson fusion and in association with a $\PW$ or $\cPZ$ boson.
It is assumed that the rate of new decays of the \PH are sufficiently small that the narrow
width approximation can be used.
The LFV \PH  decay samples are produced with  \PYTHIA 8.175~\cite{Sjostrand:pythia8}.
The background  event samples with a SM \PH are generated by \POWHEG 1.0~\cite{Nason:2004rx,Frixione:2007vw, Alioli:2010xd, Alioli:2010xa, Alioli:2008tz} with the $\Pgt$ decays modelled by \TAUOLA~\cite{TAUOLA}.
The \MADGRAPH 5.1~\cite{Alwall:2011uj}
generator is used for \cPZ+jets, \PW+jets, $\ttbar$, and diboson production, and \POWHEG for single top-quark production. The \POWHEG and \MADGRAPH generators are interfaced with \PYTHIA for parton shower and fragmentation.

\section{Event reconstruction}
A particle-flow (PF)  algorithm~\cite{CMS-PAS-PFT-09-001, CMS-PAS-PFT-10-001} combines  the information from all CMS sub-detectors to
identify and reconstruct the individual particles emerging from all
vertices: charged hadrons, neutral hadrons, photons, muons, and electrons.
These particles are then used to reconstruct  jets,
hadronic $\Pgt$ decays, and to quantify the isolation of leptons and
photons. The missing transverse energy vector is the negative vector sum of all particle
transverse momenta and its magnitude is referred to as \ETmiss. The
variable $\Delta R = \sqrt {\smash[b]{(\Delta\eta)^2 +(\Delta\phi)^2}}$ is used to
measure the separation between reconstructed objects in the detector,
where $\phi$ is the azimuthal angle (in radians) of the trajectory of the object in the
plane transverse to the direction of the proton beams.

The large number of proton interactions occurring per LHC bunch
crossing  (pileup), with an average of 21 in 2012, makes
the identification of the vertex  corresponding to the hard-scattering
process nontrivial. This affects most of the object reconstruction
algorithms: jets, lepton isolation, etc.  The tracking system is able to
separate collision vertices as close as 0.5\mm along the beam
direction~\cite{IEEE_DetAnnealing}. For each vertex, the sum of
the $\pt^2$ of all tracks associated with the vertex is computed. The
vertex for which this quantity is the largest is assumed to correspond
to the hard-scattering process, and is referred to as the primary vertex
in the event reconstruction.

Muons are reconstructed using two algorithms~\cite{Chatrchyan:2012xi}: one in which
tracks in the silicon tracker are matched to signals in the muon
detectors, and another in which a global track fit is performed, seeded by
signals in the muon systems.  The muon
candidates used in the analysis are required to be successfully reconstructed
by both algorithms.  Further identification criteria  are
imposed on the muon candidates to reduce the fraction
of tracks misidentified as muons. These include the number of measurements in the tracker and
in the muon systems, the fit quality of the global muon track and its consistency with the primary
vertex.

Electron reconstruction requires the matching
of an energy cluster in the ECAL with a track in the silicon
tracker~\cite{CMS-PAS-EGM-10-004,Khachatryan:2015hwa}. Identification criteria based on the ECAL
shower shape, matching between the track and the ECAL cluster, and consistency with the
primary vertex are imposed. Electron identification relies on a
multivariate technique that combines observables sensitive to the
amount of bremsstrahlung along the electron trajectory, the
geometrical and momentum matching between the electron trajectory and
associated clusters, as well as shower-shape observables. Additional requirements are imposed to remove electrons produced by photon conversions.

Jets  are reconstructed from all the PF objects using the anti \kt jet clustering
algorithm~\cite{Cacciari:2008gp} implemented in \textsc{FastJet}~\cite{Cacciari:fastjet}, with a distance parameter of 0.5. The jet energy
is corrected for the contribution of particles created in pileup
interactions and in the underlying event. Particles from different pileup
vertices can be clustered into a pileup jet, or significantly overlap a
jet from the primary vertex below the \PT threshold applied in the analysis.
Such jets are identified and removed~\cite{CMS-PAS-JME-13-005}.

Hadronically decaying $\Pgt$ leptons are reconstructed and
identified using the hadron plus strips (HPS) algorithm~\cite{Chatrchyan:2012zz}
which targets the main decay modes by selecting PF candidates with
one charged hadron and up to two neutral pions, or with three charged
hadrons. A photon from a neutral-pion decay can convert in the tracker
material into an electron and a positron, which can then radiate
bremsstrahlung photons. These particles give rise to several ECAL energy
deposits at the same $\eta$ value and separated in azimuthal angle,
and are reconstructed as several photons by the PF algorithm. To increase
the acceptance for such converted photons, the neutral pions are
identified by clustering the reconstructed photons in narrow strips along
the azimuthal direction.

\section{Event selection}

The event selection consists of three steps. First, a loose selection defining the basic
signature is applied. The sample is then divided into categories, according to the number of jets in the event.
Finally, requirements are placed on  a set of kinematic variables designed to suppress the backgrounds.

The loose selection for the  $\PH \to \Pgm \Pgt_{\Pe}$  channel requires an isolated  $\Pgm$
($\pt > 25$\GeV, $\abs{\eta} <2.1$) and an  isolated  $\Pe$ ($\pt > 10$\GeV, $\abs{\eta} <2.3$)  of opposite charge lying within a region of the detector that allows good identification. The $\Pe$ and $\Pgm$ are required to be separated by $\Delta R >0.1$. The
$\PH \to \Pgm \tauh$ channel requires an isolated $\Pgm$ ($\pt > 30$\GeV, $\abs{\eta} <2.1$) and an  isolated  hadronically decaying  $\Pgt$ ($\pt >30$\GeV, $\abs{\eta} <2.3$) of opposite charge. Leptons
are also required to be isolated from any jet in the event with $\pt >30$\GeV by $\Delta R > 0.4$ and to have an impact parameter consistent with the primary vertex.

The events are then divided into categories within each channel according to the number of jets in the
event. Jets are required to pass  identification criteria~\cite{CMS-PAS-JME-13-005}, have $\pt> 30$\GeV and
lie within the range $\abs{\eta} < 4.7$. The zero jet category contains signal events predominantly produced by gluon-gluon fusion.
The one-jet category contains signal events predominantly produced by gluon-gluon fusion and  a negligibly small number of events produced in association with
a W or Z boson decaying hadronically. The two jet category is enriched with signal events produced by vector boson fusion.

\begin{table}[hbtp]
 \centering
 \topcaption{Selection criteria for the kinematic variables after the loose selection.}
  \label{tab:kinematicselection}
   \cmsTable{\columnwidth}{\begin{tabular}{lccc|ccc} \hline
Variable &\multicolumn{3}{c|}{$\PH\to\Pgm\Pgt_{\Pe}$ }                 &     \multicolumn{3}{c}{$\PH \to \Pgm \tauh$}
 \\ \cline{2-7}
      [\GeVns{}]                                   &  0-jet        & 1-jet       & 2-jet         &  0-jet         & 1-jet       & 2-jet  \\ \hline
$\pt^{\Pgm}>$                                &     50        &   45        &   25          &  45            & 35          &  30    \\
$\pt^{\Pe}>$                                  &     10        &   10        &   10          &   \NA            &  \NA          &  \NA      \\
$\pt^{\tauh}>$                               &     \NA         &    \NA        &    \NA          &  35            & 40          &  40    \\
$\MT^{\Pe}<$                                   &    65         &   65        &   25          &    \NA           &   \NA         &  \NA      \\
$\MT^{\Pgm}>$                                 &    50         &   40        &   15          &    \NA           &   \NA         &  \NA      \\
$\MT^{\tauh}<$                                &     \NA         &    \NA        &    \NA          &  50            & 35          &   35   \\   \hline
      [radians]                               &                     &                \\  \hline
$\Delta \phi_{\ptvec^{\Pgm}-\ptvec^{\tauh}}>$   &     \NA         &    \NA        &    \NA          &  2.7           &  \NA          &  \NA      \\
$\Delta \phi_{\ptvec^{\Pe}-\VEtmiss}<$             &    0.5        &   0.5       &   0.3         &    \NA           &  \NA          &  \NA      \\
$\Delta \phi_{\ptvec^{\Pe}-\ptvec^{\Pgm}}>$            &    2.7        &   1.0       &    \NA          &    \NA           &   \NA         &  \NA      \\  \hline

  \end{tabular}
}
\end{table}

\begin{figure*}[hbtp]\centering
 \includegraphics[width=0.42\textwidth]{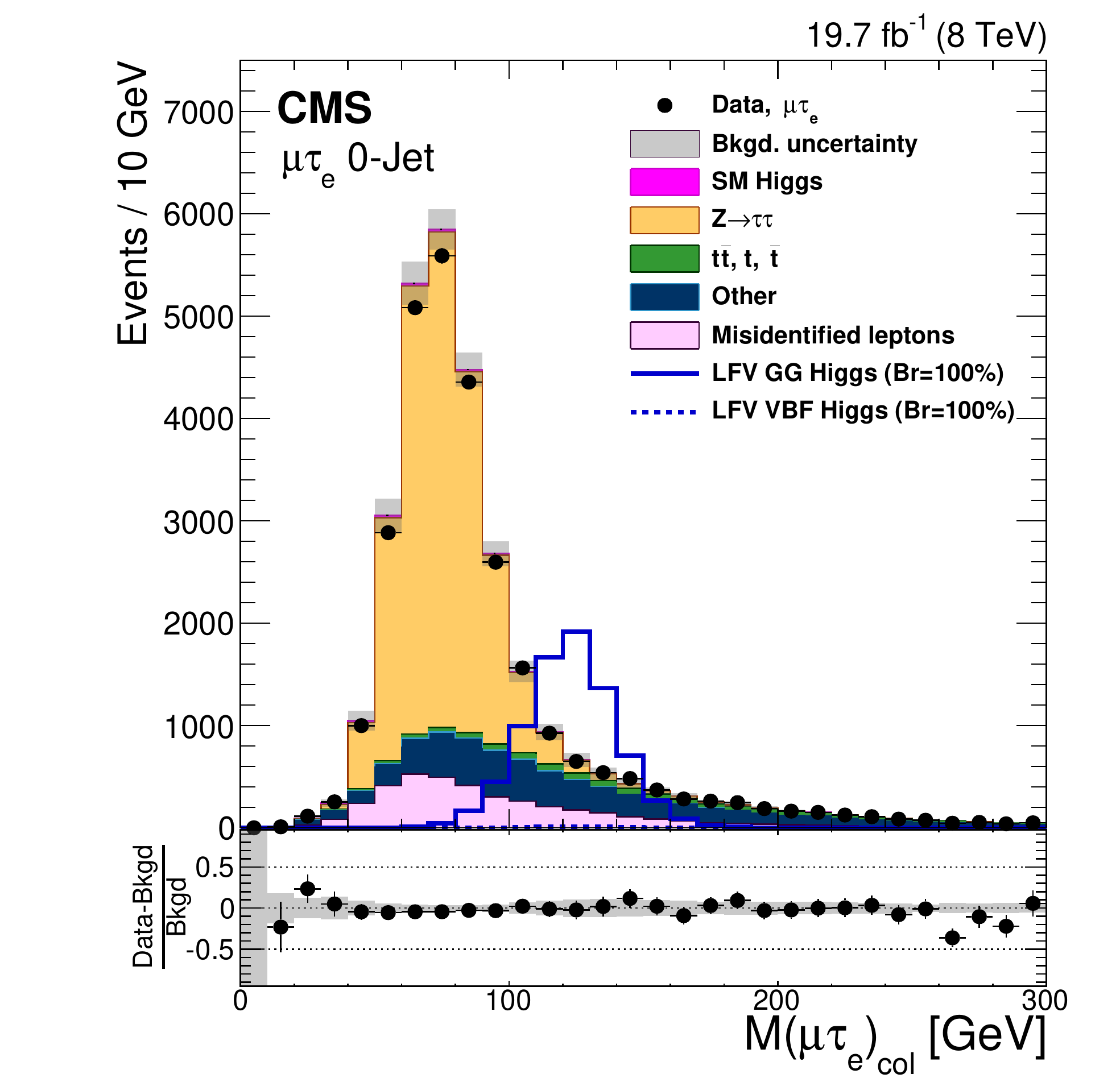}
 \includegraphics[width=0.42\textwidth]{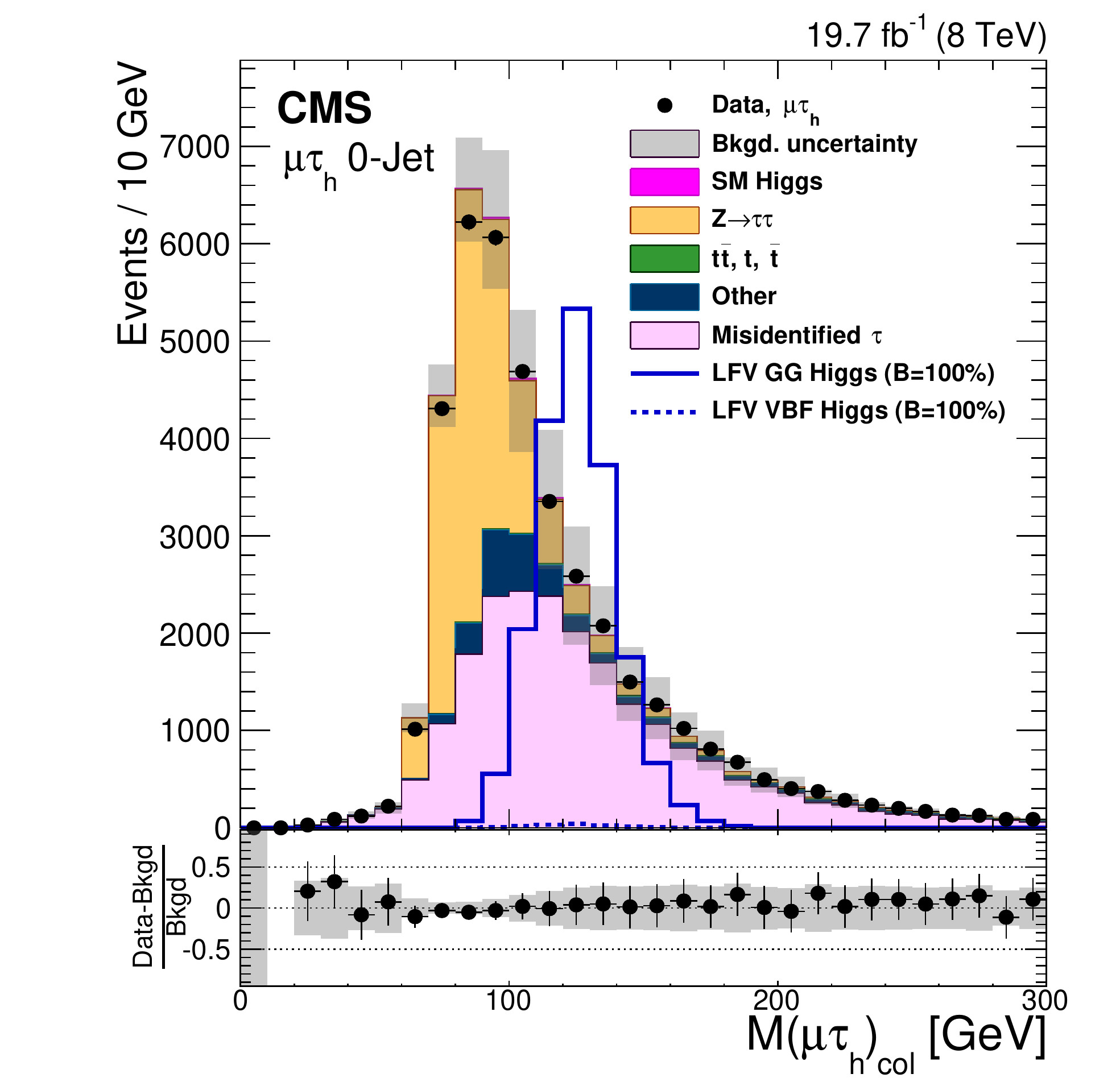}
 \includegraphics[width=0.42\textwidth]{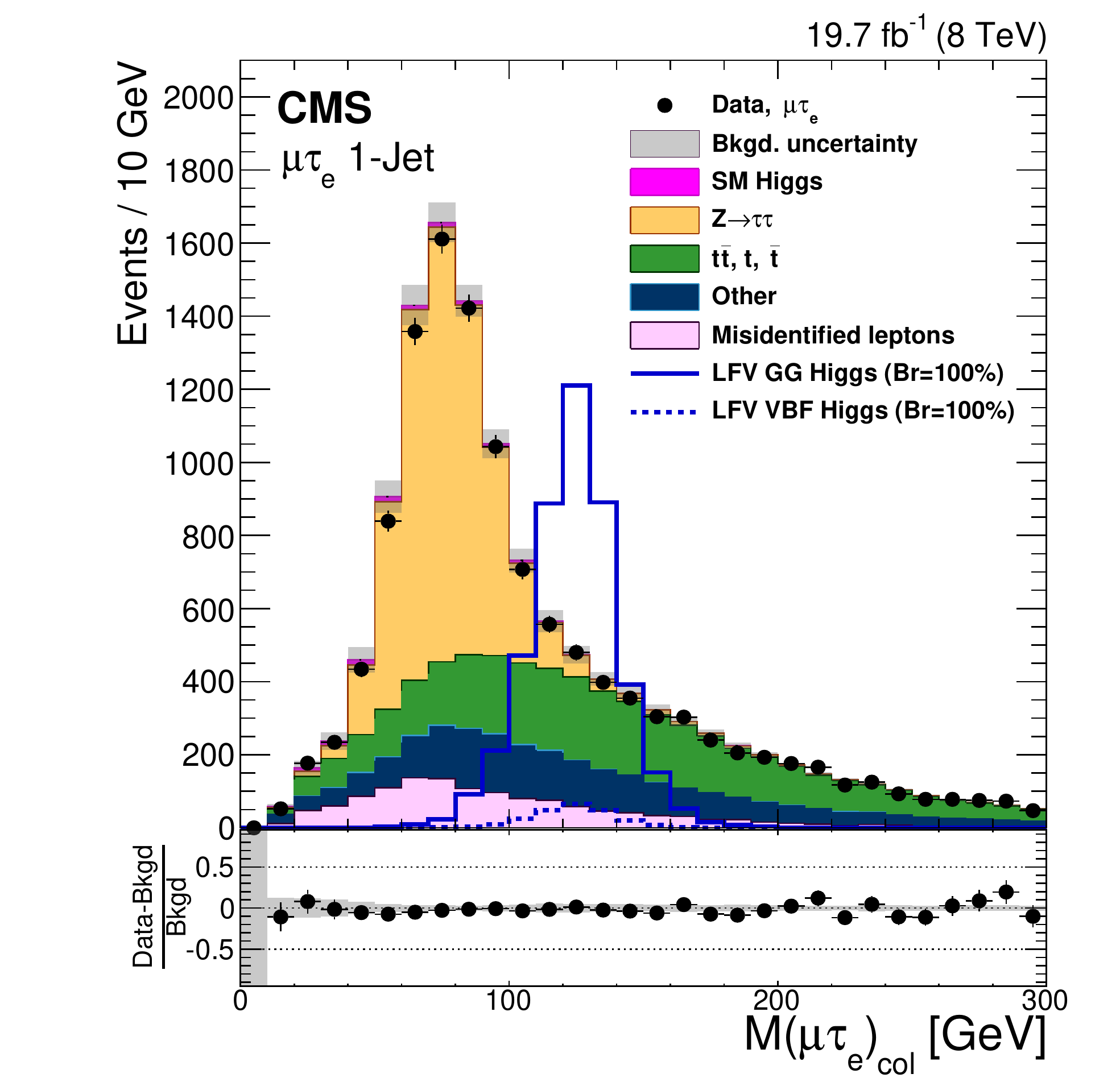}
 \includegraphics[width=0.42\textwidth]{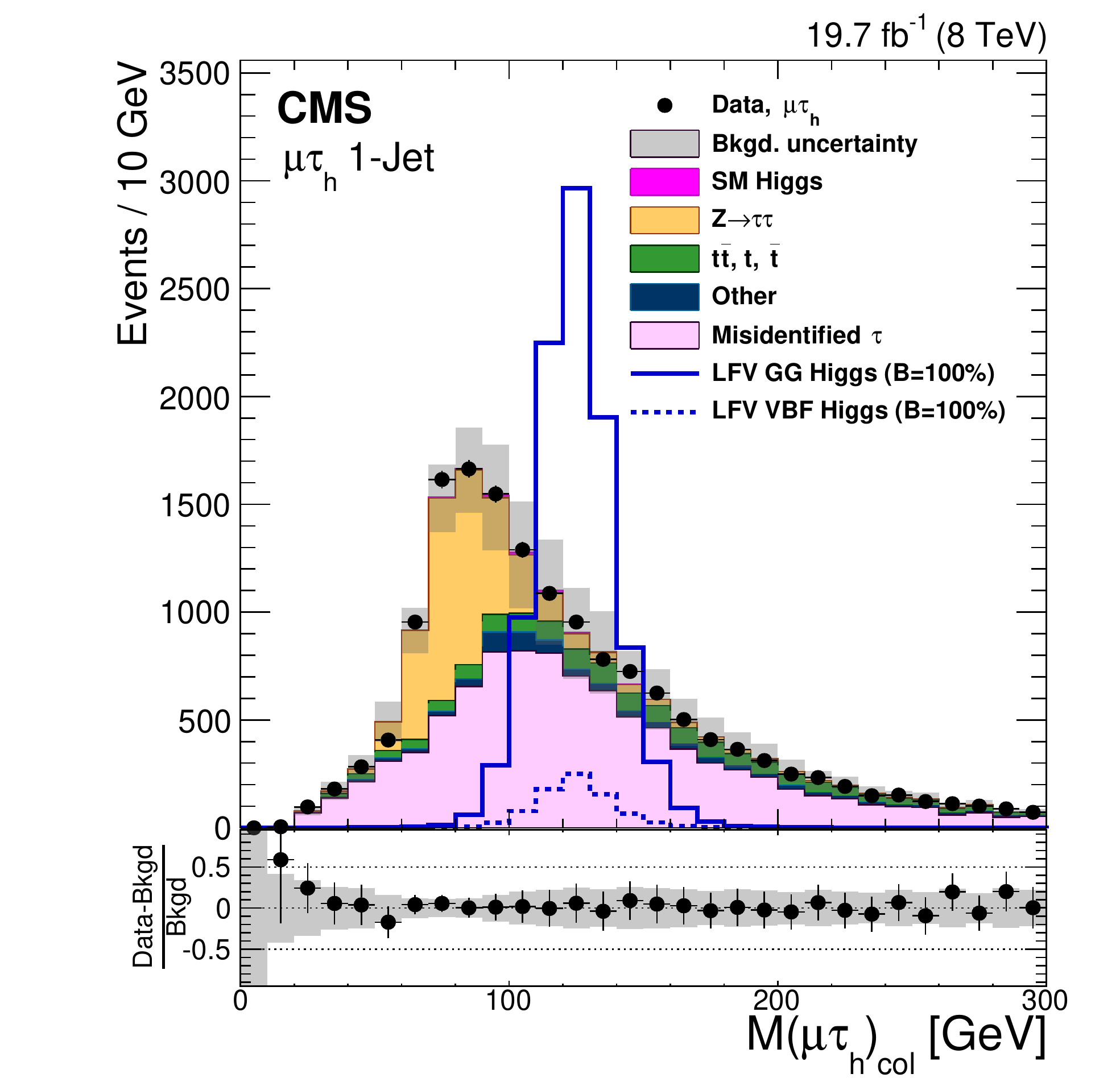}
 \includegraphics[width=0.42\textwidth]{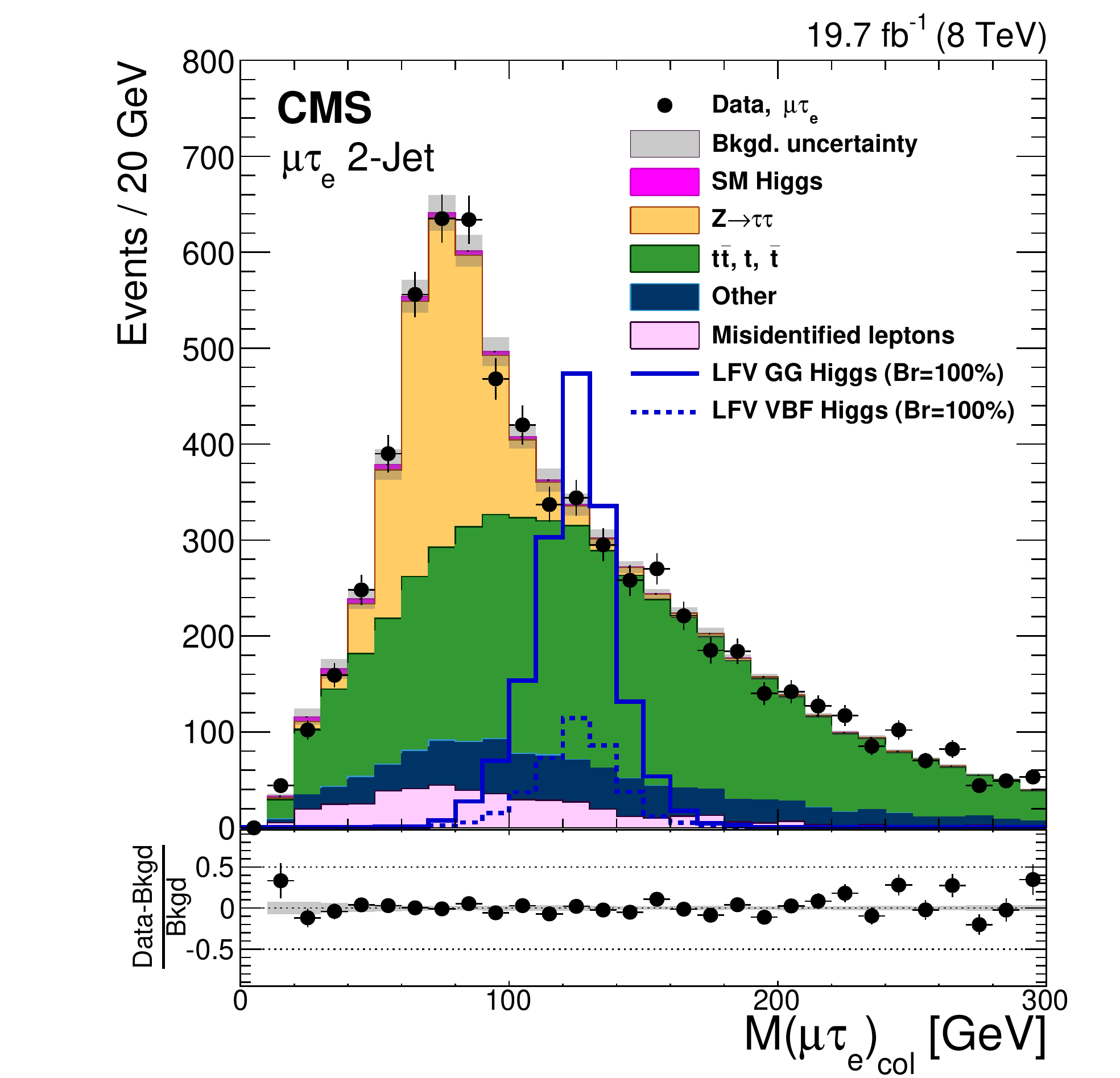}
 \includegraphics[width=0.42\textwidth]{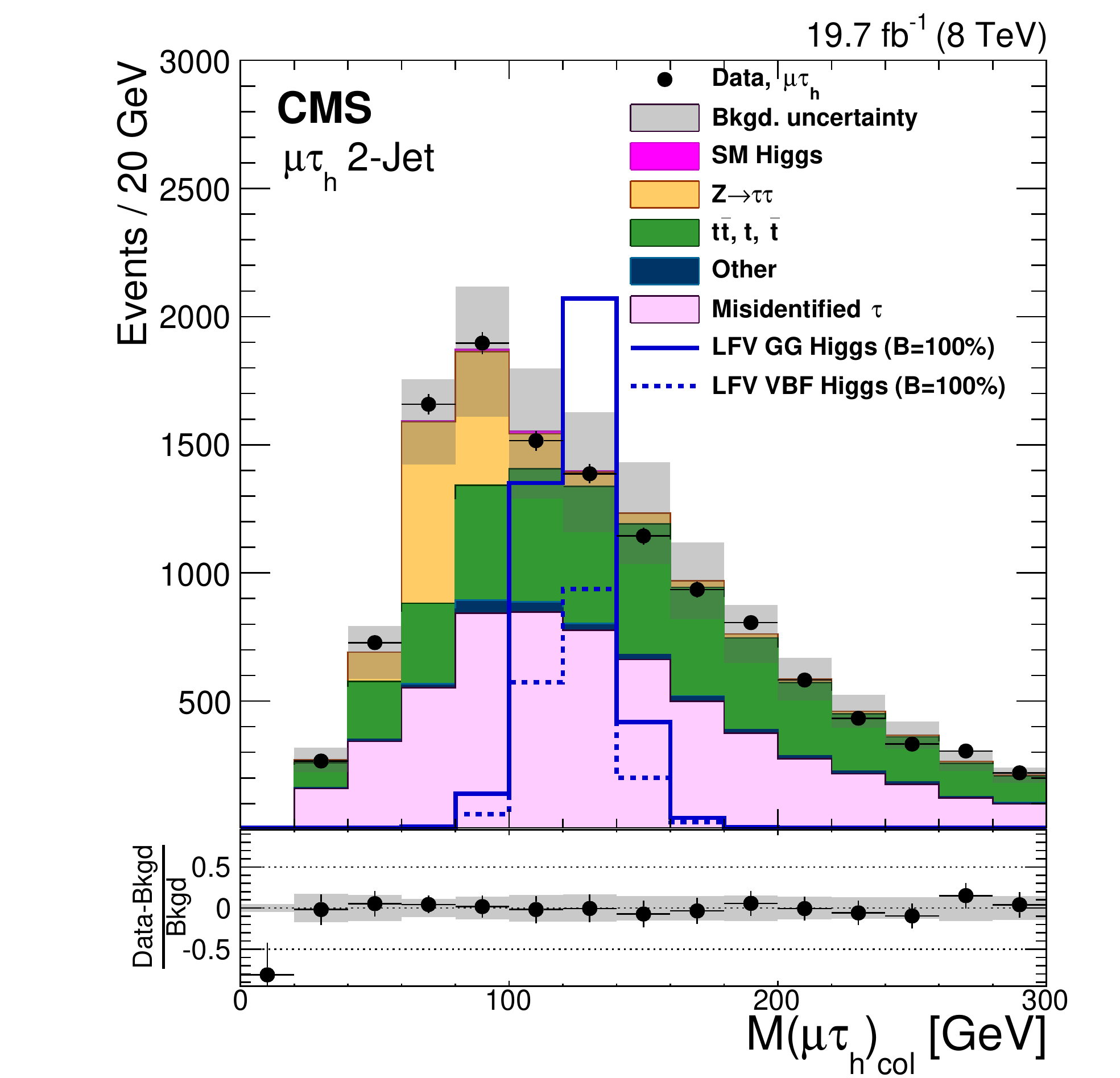}
 \caption{Distributions of the collinear mass $M_\text{col}$ for signal with $\mathcal{B}(\PH \to \Pgm \Pgt )=100\%$ for clarity, and background processes after the loose selection requirements for the LFV $\PH \to \Pgm \Pgt$ candidates for the different channels and categories compared to data. The shaded grey bands indicate the total uncertainty. The bottom panel in each plot shows the fractional difference between the observed data and the total estimated background.  Top left: $\PH \to \Pgm \Pgt_{\Pe}$ 0-jet; top right: $\PH \to \Pgm \tauh$ 0-jet;  middle left: $\PH \to \Pgm \Pgt_{\Pe}$ 1-jet; middle right: $\PH \to \Pgm \tauh$
1-jet; bottom left: $\PH \to \Pgm \Pgt_{\Pe}$ 2-jet; bottom right $\PH \to \Pgm \tauh$ 2-jet. }
 \label{fig:Mcol_after_presel_WITHDATA}\end{figure*}

The main variable for the discrimination between the signal and background is the collinear mass, $M_\text{col}$, which provides an estimator of the reconstructed \PH mass using the observed
decay products. This is constructed using the collinear approximation~\cite{Ellis:1987xu} which is based on
the observation that since the mass of the \PH  is much greater than the mass of the $\Pgt$, the $\Pgt$ decay products are highly
Lorentz boosted in the direction of the  $\Pgt$. The neutrino momenta  can be approximated to be in the same  direction as
the other visible decay products of the $\Pgt$ and the component of the missing transverse energy in the transverse direction of the
visible $\Pgt$ decay products is used to estimate the transverse component of the neutrino momentum.
Figure~\ref{fig:Mcol_after_presel_WITHDATA} shows $M_\text{col}$ distribution for the signal and background compared to data for each of the
categories in each channel after the loose selection. The simulated signal  for $\mathcal{B}(\PH \to \Pgm \Pgt )=100\%$ is shown. The principal backgrounds are estimated with data using techniques described in
Section~\ref{sec:backgrounds}.  There is good agreement between data and the background estimation. The agreement is similar  in all of the kinematic variables that are subsequently used to suppress backgrounds. The analysis is performed ``blinded'' in the region $100 < M_\text{col} < 150\GeV$.

Next, a set of kinematic variables is defined and the  criteria for  selection are determined by optimizing for $\mathrm{S}/\sqrt{\mathrm{S}+\mathrm{B}}$ where S and B are the expected signal and background event yields in the mass window $100  < M_\text{col} <  150\GeV$.
The signal event yield corresponds to the SM \PH production  cross section at $\MH=125$\GeV with
$\mathcal{B}(\PH \to \Pgm \Pgt )=10\%$. This value for the
LFV \PH branching fraction is chosen because it corresponds to the limit from indirect measurements as described in Ref.~\cite{Harnik:2012pb}. The optimization was also performed
assuming  $\mathcal{B}(\PH \to \Pgm \Pgt )=1\%$ and negligible change in the optimal values of selection criteria was observed.
The criteria for each category, and in each channel, are given in Table~\ref{tab:kinematicselection}.
The variables used are the lepton transverse momenta $\pt^{\ell}$ with $\ell=\tauh,\Pgm,\Pe$; azimuthal angles between the leptons
$\Delta \phi_{\ptvec^{\ell_{1}}-\ptvec^{\ell_{2}}}$;
azimuthal angle $\Delta \phi_{\ptvec^{\ell}-\VEtmiss}$; the transverse mass
$\MT^{\ell}=\sqrt{\smash[b]{2\pt^{\ell}\ETmiss(1-\cos{\Delta \phi_{\ptvec^{\ell}-\VEtmiss}})}}$.
Events  in  the 2-jet
category are required to have two jets separated by a pseudorapidity gap ($\abs{\Delta \eta} > 3.5$)  and to have a
dijet invariant  mass greater than 550\GeV. In the $\PH \to \Pgm \Pgt_{\Pe}$ channel events in which at least one of the jets  identified
as coming from a b-quark decay are  using the combined secondary-vertex b-tagging
algorithm~\cite{Chatrchyan:2012jua} are vetoed, to suppress backgrounds from top quark decays.

\section{Background Processes}
\label{sec:backgrounds}
The contributions of the dominant background processes are estimated with data  while less significant
backgrounds are estimated using simulation.  The largest backgrounds come from $\cPZ \to \Pgt \Pgt$
and from misidentified leptons in \PW+jets and QCD multijet production.

\subsection{\texorpdfstring{$ \cPZ \to \Pgt \Pgt$}{Z to tau tau}}
\label{sec:hmuepftauembed}
\label{sec:hmueembed}

The  $\cPZ \to \Pgt \Pgt$  background contribution is estimated using an embedding technique~\cite{CMS:2011aa,Chatrchyan:2014nva} as follows.
A sample of $\cPZ \to \Pgm \Pgm$ events is taken from data using a loose $\Pgm$ selection. The
two muons are then replaced
with PF particles resulting from the reconstruction of simulated $\Pgt$ lepton decays. Thus, the key features of the event topology such as the
jets, missing transverse energy and underlying event are taken directly from data with only the $\Pgt$ decays being simulated. The normalization
of the sample is obtained from the simulation. The technique is validated by comparing the $\Pgt$ lepton identification efficiencies
estimated with an embedded decay sample, using simulated $\cPZ \to \Pgm\Pgm$ events, to those from  simulated $\cPZ \to \Pgt \Pgt$ decays.

\subsection{Misidentified leptons}
\label{sec:fakes}

Leptons can arise from misidentified PF objects in \PW+jets and QCD multijet processes.
This background is estimated with data. A sample with similar kinematic
properties to the signal sample but enriched in \PW+jets and QCD multijets
is defined. Then the probability for PF objects to be misidentified as leptons is measured
in an independent data set, and this probability is applied to the enriched sample to
compute the misidentified lepton background in the signal region.
The technique is shown schematically in Table~\ref{tab:fakeratediagram} in which four regions
are defined including the signal and background enriched regions and two control regions used for validation
of the technique. It is employed slightly differently in the
$\PH \to \Pgm \Pgt_{\Pe}$ and $\PH \to \Pgm \tauh$ channels.
The lepton isolation requirements used to define the enriched regions in each channel
are slightly different.

In the $\PH \to \Pgm \Pgt_{\Pe}$ channel, region I is the signal region in which an isolated $\Pgm$ and an isolated $\Pe$ are required.
Region III is a data sample in which all the analysis selection criteria are applied except that
one of the leptons is required to be not-isolated. Thus, there are two components: events with an
isolated $\Pgm$ and not-isolated $\Pe$ events, as well as events with an isolated $\Pe$ and not-isolated $\Pgm$ events. There is negligible number of signal events in region III. Regions II and IV are data samples formed with the same selection criteria as regions I and III, respectively, but with same-sign rather than opposite-sign leptons. The kinematic distributions of the same-sign samples are very similar to the opposite-sign samples

\begin{figure*}[hbtp]\centering
\includegraphics[width=0.48\textwidth]{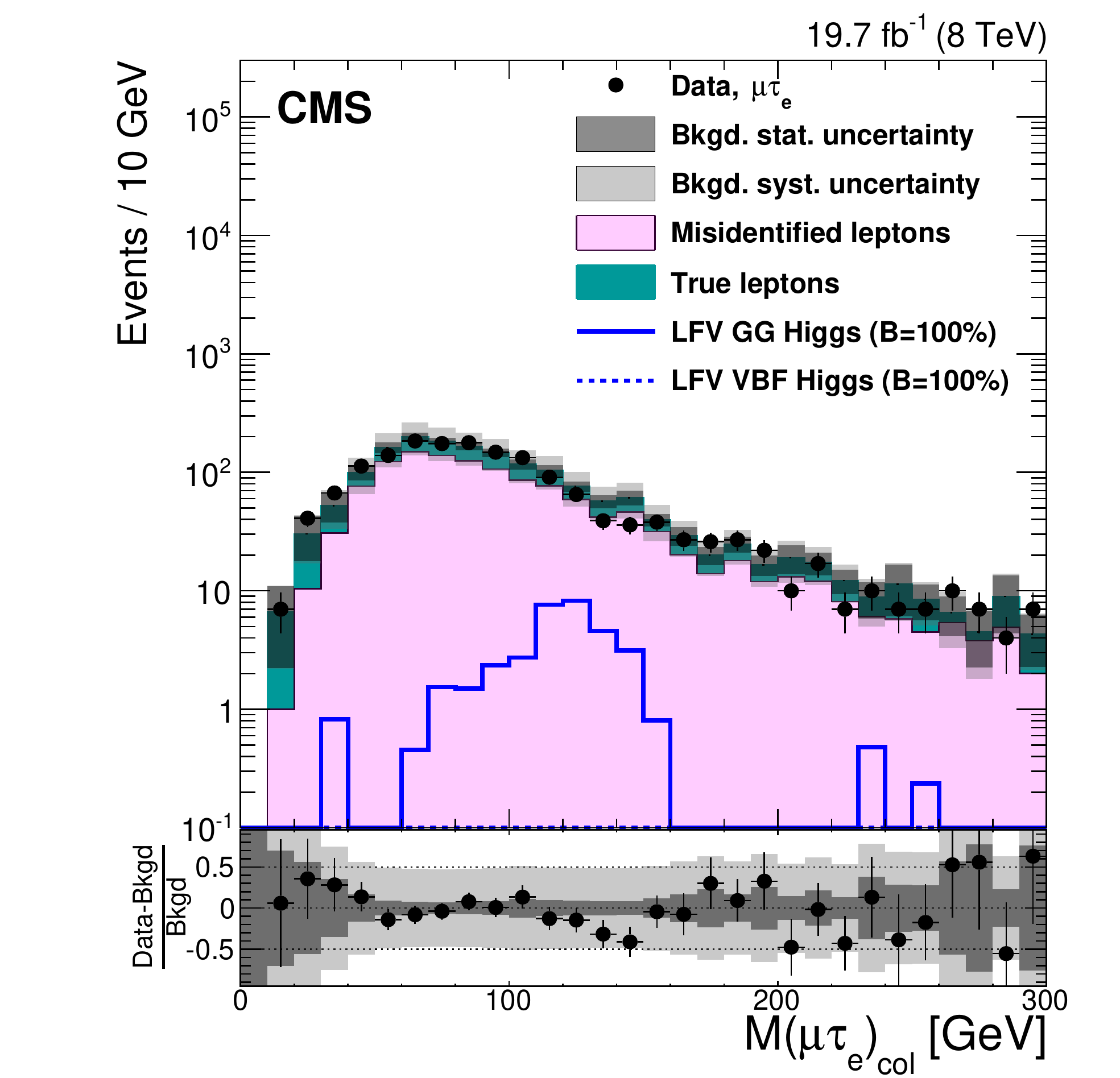}
\includegraphics[width=0.48\textwidth]{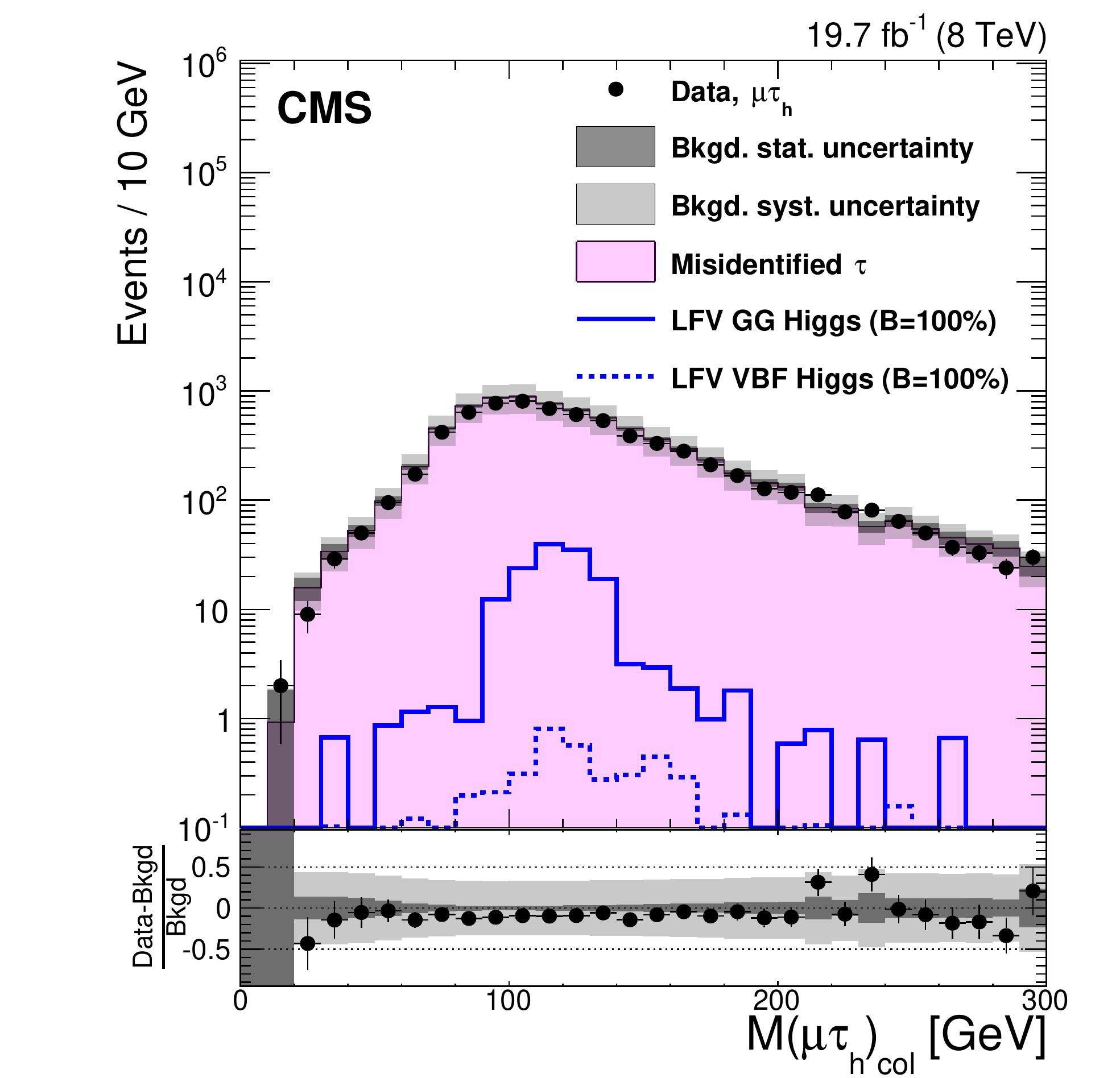}
\caption{Distributions of $M_\text{col}$ for region II compared to the estimate
from scaling the region IV sample by the measured misidentification rates. The bottom panel in each plot shows the fractional difference between the observed data and the estimate. Left:  $\PH \to \Pgm \Pgt_{\Pe}$. Right: $\PH \to \Pgm \tauh$. }
\label{fig:samesign_fakes}\end{figure*}

\begin{table}[hbt]
 \centering
 {
 \renewcommand{\arraystretch}{1.1}
 \topcaption{Schematic to illustrate the application of the method used to estimate the misidentified lepton ($\ell$) background. Samples
are defined by the charge of the two leptons and by the isolation requirements on each. Charged conjugates are assumed.}
  \label{tab:fakeratediagram}

  \begin{tabular}{c|c}
\rule[-5pt]{0pt}{20pt}
\textbf{Region I}              &  \textbf{Region II}             \\
$\ell^{+}_{1}$(isolated)  &  $\ell^{+}_{1}$(isolated)             \\
$\ell^{-}_{2}$(isolated)  &  $\ell^{+}_{2}$(isolated)             \\

\hline
\rule[-5pt]{0pt}{20pt}
\textbf{Region III}           &  \textbf{Region IV}             \\
$\ell^{+}_{1}$(isolated)  &  $\ell^{+}_{1}$(isolated)             \\
$\ell^{-}_{2}$(not-isolated )  &  $\ell^{+}_{2}$(not-isolated)             \\
  \end{tabular}
}

\end{table}
The sample in region III is dominated by \PW+jets and QCD multijets but with small
contributions from $\PW\PW,\cPZ\cPZ$ and $\PW\cPZ$  that are subtracted using
simulation. The misidentified  $\Pgm$ background in region I is then estimated by multiplying the event yield in region III by a
factor $f_{\Pgm}\cdot\epsilon_\text{trigger}$, where $f_{\Pgm}$ is the ratio
of not-isolated to isolated $\Pgm$'s. It is computed in an independent data sample $\cPZ \to \Pgm \Pgm + X$, where $X$ is an object identified as a $\Pgm$, in bins of $\pt$ and $\eta$. The $\cPZ \to \Pgm \Pgm + X$ sample is corrected for contributions from $\PW\PW,\cPZ\cPZ$ and $\PW\cPZ$ using  simulated samples.
A correction $\epsilon_\text{trigger}$ is made to
account for the difference in trigger efficiency for selection of events with isolated $\Pe$ and  not-isolated
$\Pgm$ versus the events with isolated $\Pe$ and isolated $\Pgm$.
The misidentified $\Pe$ background is computed in exactly the same way.
The technique is validated by using the  same-sign data from regions II and IV
as shown schematically  in Table~\ref{tab:fakeratediagram}.
In Fig.~\ref{fig:samesign_fakes}(left) the observed data yield in region II is compared
to the estimate from scaling the region IV sample
by the measured misidentification rates. The region II  sample is dominated by
misidentified leptons but also includes  small contributions of true leptons arising
from vector boson decays,  estimated with simulated samples.

In the $\PH \to \Pgm \tauh$ channel, the $\tauh$ candidate can come from a  misidentified
jet with a number of sources, predominantly  $W\mathrm{+jets}$ and QCD multijets,
but also $\cPZ \to \Pgm \Pgm\text{+jets}$ and $\ttbar$. In this case the enriched background
regions are defined with $\Pgt_h$ candidates that pass a looser isolation requirement, but do not pass the
signal isolation requirement. The misidentification rate $f_{\tauh}$ is then defined as the fraction of $\tauh$ candidates with
the looser isolation that also pass the signal isolation requirement. It is measured in
observed  $\cPZ \to \Pgm \Pgm +X$ events, where $X$ is
an object identified as a $\tauh$.
The misidentification rate  measured in  $\cPZ \to \Pgm \Pgm  +X$ data is checked by comparing to that
measured in $\cPZ \to \Pgm \Pgm +X$ simulation and found to be in good agreement.
The misidentified background in the signal region (region I) is estimated by multiplying
the event yield in region III by a factor $f_{\tauh}/(1-f_{\tauh})$.
The procedure is validated with same-sign $\Pgm\Pgt$ events in the same way as
for the $\PH \to \Pgm \Pgt_{\Pe}$ channel above. Figure~\ref{fig:samesign_fakes}(right) shows
the data  in region II compared to the estimate from scaling
region IV by the misidentification rates.

The method assumes that the misidentification rate in $\cPZ \to \Pgm \Pgm +X$ events is the same as
for \PW+jets and QCD processes. To test this assumption the misidentification rates are measured in a QCD jet data control
sample. They are found to be consistent. Finally as a cross-check the study has been performed also as a function of the number of
jets in the event and similar agreement is found.

\subsection{Other backgrounds}
The SM \PH decays in the $\PH \to \Pgt\Pgt$ channel provide a small background that is
estimated with simulation. This background is suppressed by the kinematic selection criteria
and peaks below 125\GeV. The $\PW$ leptonic decay from  $\ttbar$ produces opposite-sign dileptons and \ETmiss. This background is estimated with simulated $\ttbar$
events using the shape of the $M_\text{col}$ distribution from simulation and a data control region for
normalization. The control region is the  2-jet selection but with the additional requirement that
at least one of the jets is b-tagged in order to enhance the $\ttbar$ contribution.
Other smaller backgrounds come from $\PW\PW$, $\cPZ\cPZ\text{+jets}$, $\PW\gamma\text{+jets}$ and single
top-quark production. Each of these is estimated with simulation.

\section{Systematic uncertainties}
To set upper bounds on the signal strength, or determine a signal significance, we use the
CL$_{\mathrm{s}}$ method~\cite{Junk,Read2}.  A binned likelihood is used, based on the
distributions of $M_\text{col}$ for the signal and the various background sources.
Systematic uncertainties are represented by nuisance parameters, some of which only affect
the background and signal normalizations, while  others affect the shape and/or
normalization of the $M_\text{col}$ distributions.

\subsection{Normalization uncertainties}

\begin{table*}[t]
 \centering
  \topcaption{Systematic uncertainties in the expected event yield in \%. All uncertainties are treated as correlated between the categories, except where there are two numbers. In
this case the number denoted with * is treated as uncorrelated between categories and the
total uncertainty is the sum in quadrature of the two numbers.}
  \label{tab:systematics}
{
\begin{tabular}{lccc|ccc} \hline
Systematic  uncertainty                                &  \multicolumn{3}{c|}{$\PH \to \Pgm \Pgt_{\Pe}$}& \multicolumn{3}{c}{$\PH \to \Pgm \tauh$}    \\ \cline{2-7}
                                                       &  0-Jet  & 1-Jet  & 2-Jets     &  0-Jet    & 1-Jet     & 2-Jets     \\ \hline
electron trigger/ID/isolation                          &   3   &   3  &   3     &    \NA      &   \NA       &  \NA        \\
muon  trigger/ID/isolation                             &   2   &   2  &   2     &    2    &  2      &  2      \\
hadronic tau efficiency                                &   \NA     &   \NA    &   \NA       &    9    &  9      &  9      \\
luminosity                                             &  2.6  &  2.6 &  2.6    &  2.6    &  2.6    &  2.6    \\
$\cPZ \to \Pgt \Pgt$ background                           &   3+3*&  3+5*&  3+10*  &   3+5*  &   3+5*  &   3+10* \\
$\cPZ \to \Pgm\Pgm,\Pe\Pe$ background                           &   30  &  30  &  30     &   30    &   30    &   30    \\
misidentified $\Pgm,\Pe$  background                      &  40   &  40  &  40     &    \NA      &   \NA       &   \NA       \\
misidentified $\tauh$  background                           &  \NA      &   \NA    &    \NA      &   30+10*&  30     &  30     \\
$\PW\PW,\cPZ\cPZ\text{+jets}$ background                                 &  15   &  15  &   15    &  15     &  15     &  65     \\
$\ttbar$ background                         &  10  &  10 &  10+10* &  10    &  10    &  10+33* \\
$\PW +\gamma$ background                                 &  100  &  100 &  100   &     \NA     &    \NA      &    \NA       \\
b-tagging veto                                         &    3  &   3  &   3     &    \NA      &    \NA      &    \NA       \\
single top production background                       &  10   &  10  &  10    &  10    &  10    &   10    \\ \hline
  \end{tabular}
}

\end{table*}

The uncertainties are summarized in Tables~\ref{tab:systematics} and~\ref{tab:theory_systematics}. The
uncertainties in the $\Pe$ and $\Pgm$ selection efficiency (trigger, identification and isolation) are estimated using
the ``tag and probe'' technique in $\cPZ \to \Pe\Pe,\Pgm\Pgm$ data~\cite{CMS:2011aa}. The identification
efficiency of hadronic $\Pgt$ decays is estimated using the ``tag and probe'' technique
in $\cPZ \to \Pgt \Pgt $ data~\cite{Chatrchyan:2012zz}.
The uncertainty in the
$\cPZ \to \Pgt\Pgt$ background comes predominantly from the uncertainty in the $\Pgt$ efficiency.
The uncertainties in the estimation of the misidentified lepton  rate come from the
difference in rates measured in different data samples (QCD multijets and \PW+jets).
The uncertainty in the production
cross section of the backgrounds that have been  estimated by simulation is also
included.

There are several uncertainties on the \PH production cross section,
which depend on the  production mechanism contribution and the analysis category. They are given in Table~\ref{tab:theory_systematics}.
These affect the LFV \PH and the SM \PH background equally, and are treated as 100\% correlated.
The parton distribution function (PDF) uncertainty is evaluated by comparing the yields in each category, when spanning
the  parameter range of a number of different independent PDF sets including CT10~\cite{Nadolsky:2008zw}, MSTW~\cite{Martin:2009iq},
NNPDF~\cite{Ball:2010de} as recommended by  PDF4LHC~\cite{Botje:2011sn}. The scale uncertainty is estimated by  varying the renormalization, $\mu_{R}$, and factorization scales, $\mu_{F}$,  up and down by one half or two times the nominal scale ($\mu_{R}=\mu_{F}=\MH/2$) under the constraint $0.5 < \mu_{F}/\mu_{R} < 2$~\cite{Dittmaier:2011ti}. The underlying event
and parton shower uncertainty is estimated by using two different \PYTHIA tunes, AUET2 and Z2*. Anticorrelations arise due to  migration of events between the categories and are expressed as negative numbers.

\begin{table*}[hbtp]
 \centering
  \topcaption{Theoretical uncertainties in \% for Higgs boson production. Anticorrelations arise due to  migration of events between the categories and are expressed as negative numbers. }
  \label{tab:theory_systematics}
  \begin{tabular}{lccc|ccc} \hline
Systematic uncertainty                  &  \multicolumn{3}{c|}{Gluon-Gluon Fusion} &  \multicolumn{3}{c}{Vector Boson Fusion}  \\ \cline{2-7}
                                &    0-Jets  & 1-Jets  & 2-Jets   & 0-Jet & 1-Jet  & 2-Jets  \\ \hline
parton distribution function         &    $+9.7$  &  $+9.7$ &   $+9.7$ & $+3.6$  &   $+3.6$  &  $+3.6$  \\
renormalization/factorization scale           &    $+8$    &  $+10$   &  $-30$   & $+4$     &   $+1.5$  & $+2$   \\
underlying event/parton shower  &   $+4$     & $-5$   &  $-10$   & $+10$    &   $<$1    & $-1$   \\ \hline
  \end{tabular}

\end{table*}

\subsection{\texorpdfstring{$M_\text{col}$}{M[col]} shape uncertainties}
The systematic uncertainties that lead to a change in the shape of the  $M_\text{col}$
distribution are summarized in Table~\ref{tab:shape_systematics}.
\begin{table}[hbtp]
 \centering
  \topcaption{Systematic uncertainties in \% for the shape of the signal and background templates.}
  \label{tab:shape_systematics}
  {
  \begin{tabular}{lll} \hline
Systematic uncertainty                                 &  $\PH \to \Pgm \Pgt_{\Pe}$      &   $\PH \to \Pgm \tauh$                                      \\ \hline
hadronic tau energy scale                              &   \NA         &    3                                    \\
jet energy scale                                       &   3--7     &    3--7                                       \\
unclustered energy scale                                &   10      &    10                                   \\
$\cPZ \to \Pgt\Pgt$ bias                                 &   100 &     \NA                                          \\   \hline
  \end{tabular}
}

\end{table}
In the embedded $\cPZ \to \Pgt \Pgt$ $M_\text{col}$ distribution, used to estimate the
$\cPZ \to \Pgt \Pgt$ background,  a 1\% shift has been observed with respect
to $\cPZ \to \Pgt \Pgt$ simulations by comparing the means of both  distributions.
This occurs only in the $\PH \to \Pgm \Pgt_{\Pe}$ channel.
The $M_\text{col}$ distribution  has been corrected for this effect  and a
100\% uncertainty on this shift is used as a systematic uncertainty for the possible bias.
The jet energy scale has been studied extensively and a standard
prescription for corrections~\cite{CMS-PAPERS-JME-10-011} is used in all CMS analyses.
The overall scale is set using
$\gamma$+jets events and the most significant uncertainty arises from  the photon energy scale.
A number of other uncertainties such as jet fragmentation modeling, single pion response
and uncertainties in the pileup corrections are also included. The jet energy scale
uncertainties (3--7\%) are applied as a function of
$\pt$ and $\eta$, including all correlations, to all jets in the event, propagated to the
missing energy, and the resultant  $M_\text{col}$ distribution is used in the fit.
There is also an additional uncertainty to account for the unclustered energy scale uncertainty. The unclustered energy comes from jets below
10\GeV and PF candidates not within jets. It is also propagated to the missing transverse energy.
These effects cause a  shift of the $M_\text{col}$ distribution.
The $\Pgt$ energy scale is estimated by comparing $\cPZ \to \Pgt \Pgt $ events in data and simulation.
An uncertainty of 3\% is derived from this comparison.
The uncertainty is applied by  shifting  the $\pt$ of the $\Pgt$ candidates in the event and using the
resultant $M_\text{col}$ distribution in the fit.
Finally, the $M_\text{col}$ distributions used in the fit  have a statistical uncertainty in each mass bin that is included as an uncertainty which is  uncorrelated between the bins.

Potential uncertainties in the shape of the misidentified lepton backgrounds have also been considered. In the \mbox{$\PH \to \Pgm \Pgt_{\Pe}$} channel the misidentified lepton rates
$f_{\Pgm},f_{\Pe}$ are measured and applied in bins of lepton $\pt$ and $\eta$. These rates are
all adjusted up or down by one standard deviation ($\sigma$)  and the differences in the shape of
the resultant $M_\text{col}$ distributions  are then used as nuisance parameters in the fit.
In the $\PH \to \Pgm \tauh$ channel the $\Pgt$ misidentification rate $f_{\Pgt}$ is found to be approximately flat in $\pt$ and $\eta$. To
estimate the systematic uncertainty the $\pt$ distribution of $f_{\Pgt}$ is fit with a linear function and
the rate recomputed from the fitted slope and intercept. The modified $M_\text{col}$ distribution that results
from the recomputed background is then used to evaluate the systematic uncertainty.

\begin{figure*}[hbtp]\centering
 \includegraphics[width=0.48\textwidth]{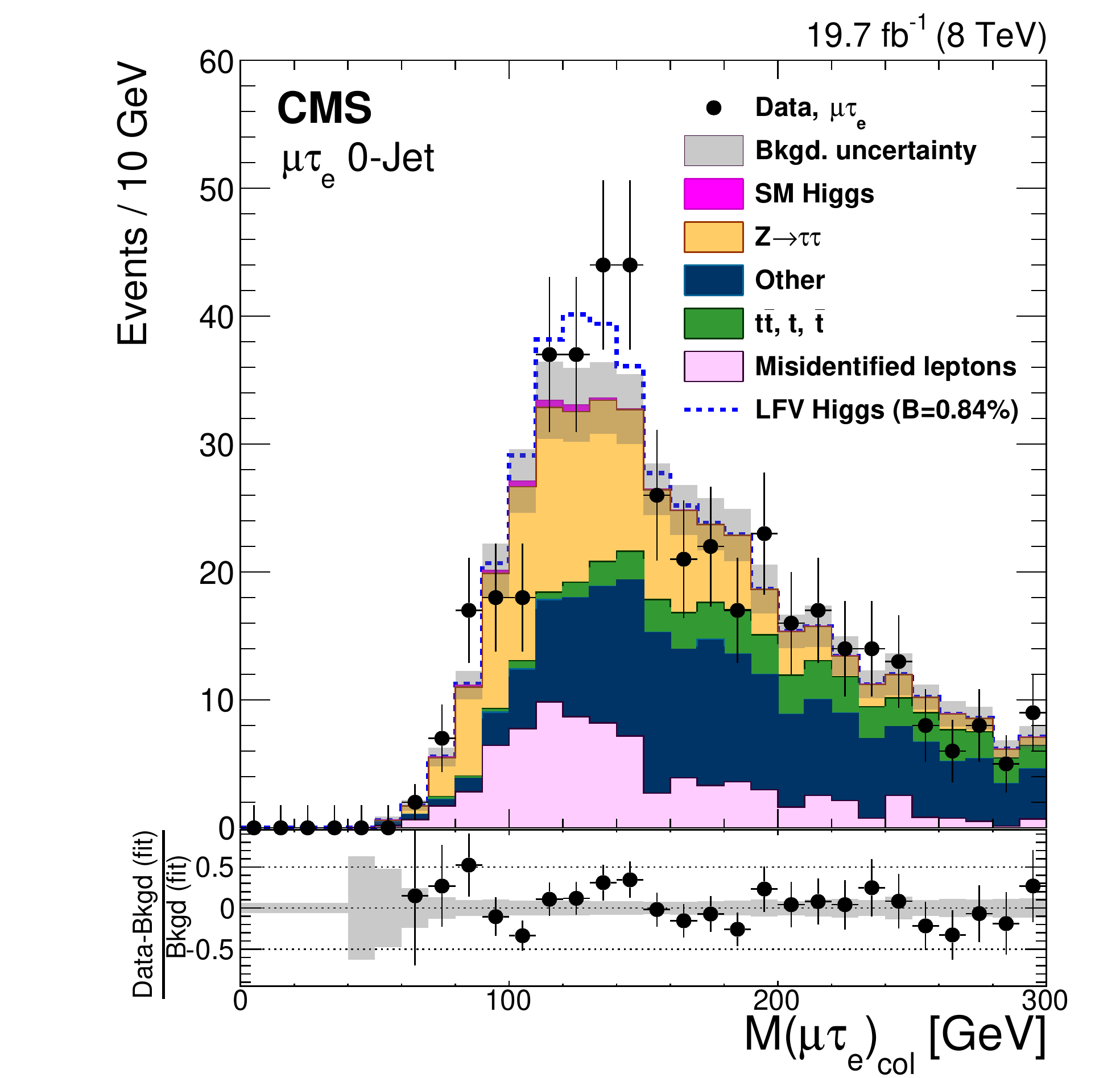}
 \includegraphics[width=0.48\textwidth]{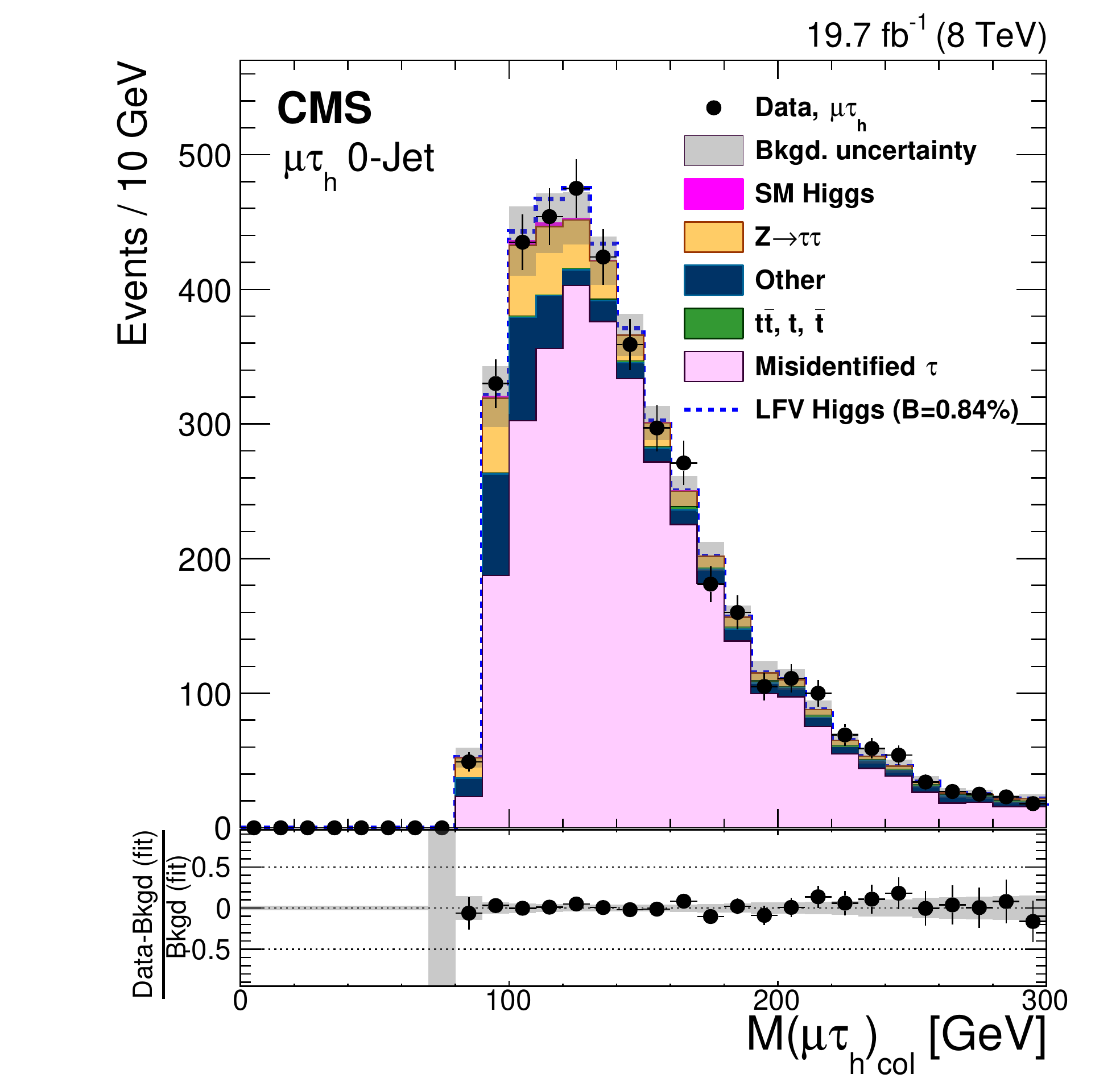}
 \includegraphics[width=0.48\textwidth]{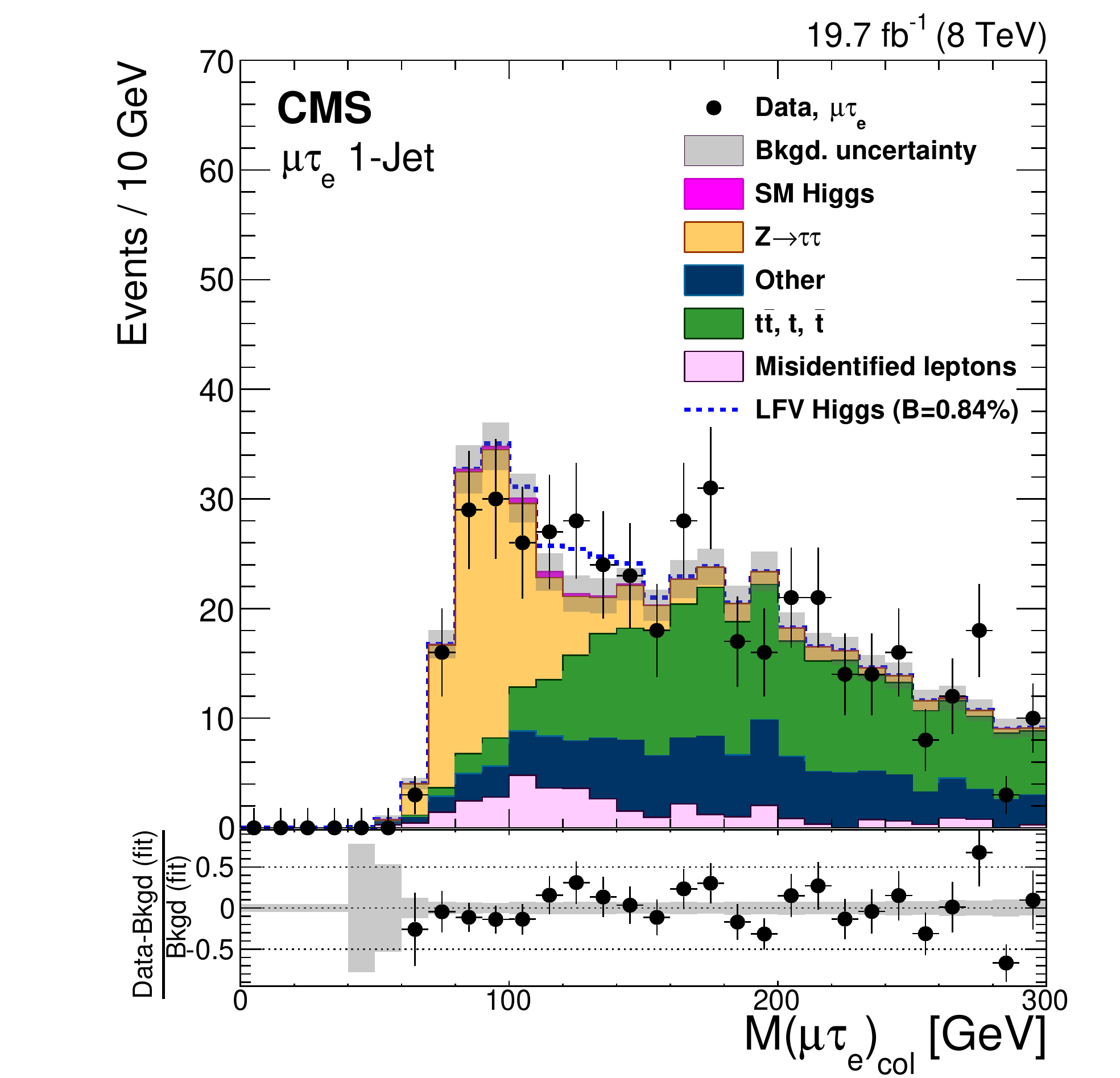}
 \includegraphics[width=0.48\textwidth]{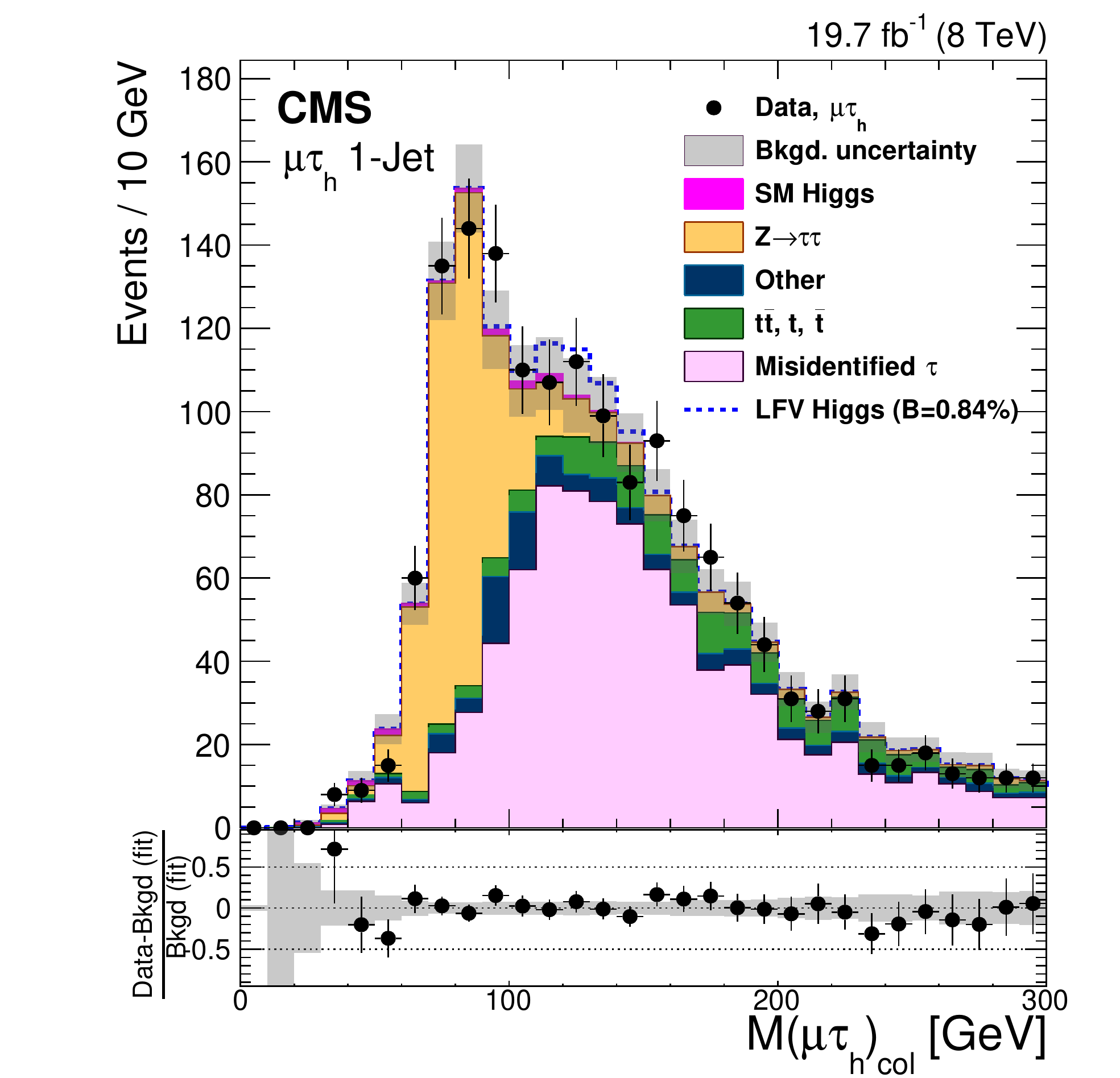}
 \includegraphics[width=0.48\textwidth]{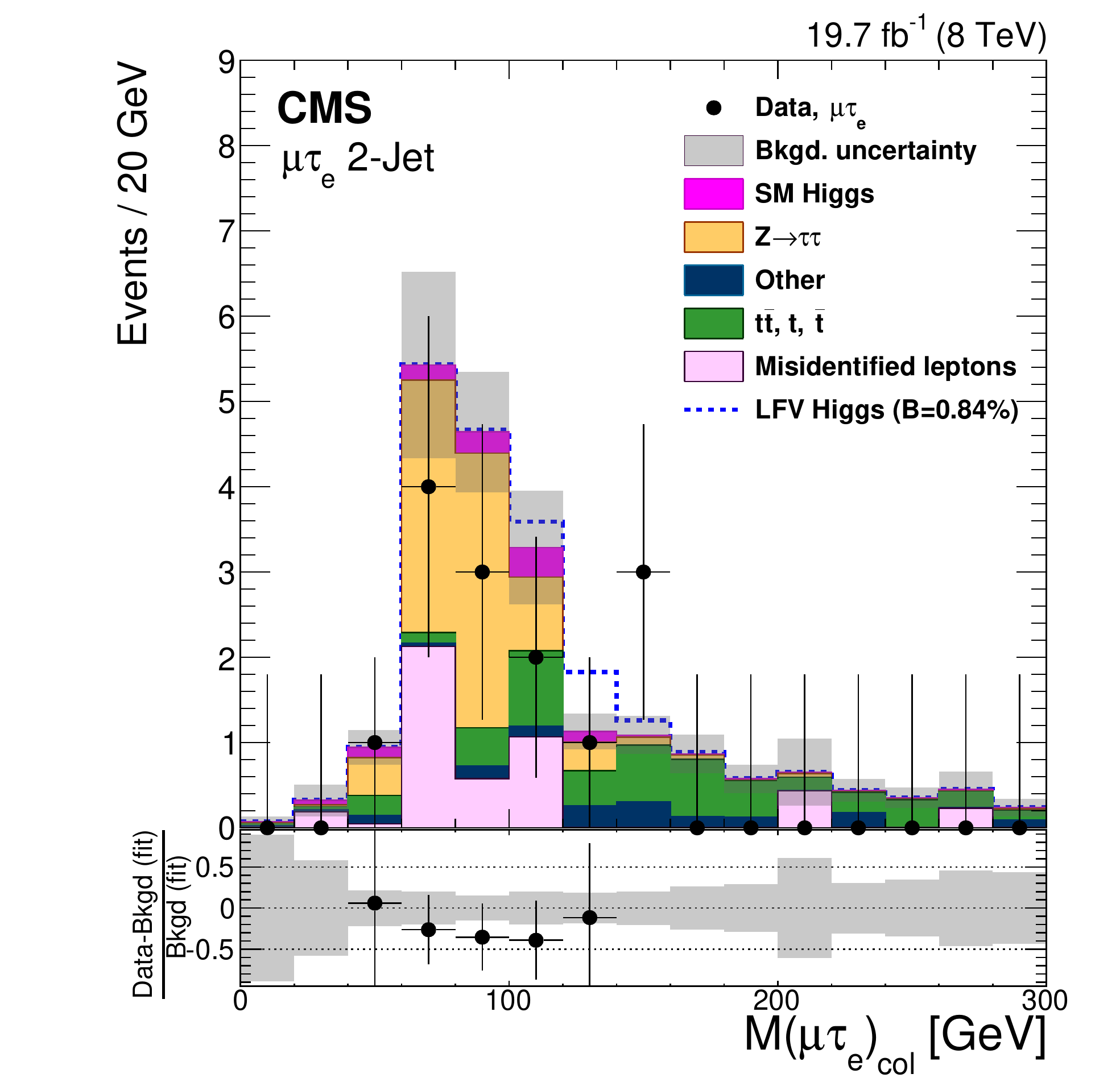}
 \includegraphics[width=0.48\textwidth]{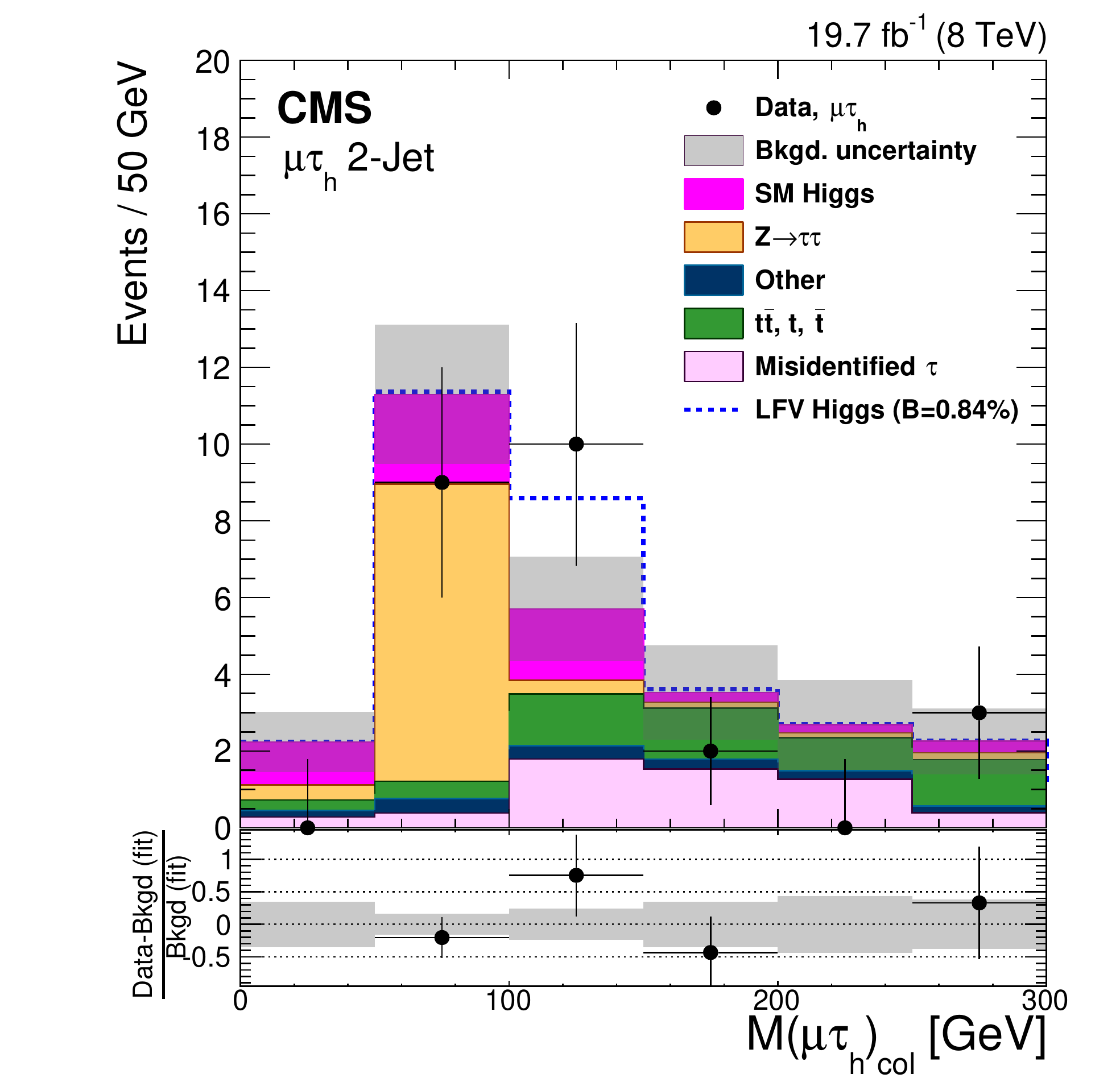}
 \caption{Distributions of the collinear mass $M_\text{col}$ after fitting for signal and background  for the LFV $\PH \to \Pgm \Pgt$ candidates in
the different
channels and categories compared to data.
The distribution of the simulated LFV Higgs boson sample is shown for the best fit branching fraction
of $\mathcal{B}(\PH \to \Pgm \Pgt )=0.84\%$.
The bottom panel in each plot shows the fractional difference between the observed data and the fitted background. Top left: $\PH \to \Pgm \Pgt_{\Pe}$ 0-jet; top right: $\PH \to \Pgm \tauh$ 0-jet;
middle left: $\PH \to \Pgm \Pgt_{\Pe}$ 1-jet; middle right: $\PH \to \Pgm \tauh$ 1-jet; bottom left: $\PH \to \Pgm \Pgt_{\Pe}$ 2-jet;
bottom right $\PH \to \Pgm \tauh$ 2-jet.}
 \label{fig:Mcol_Postfit}\end{figure*}

\section{Results}

The $M_\text{col}$ distributions after the fit for signal and background contributions are shown in Fig.~\ref{fig:Mcol_Postfit} and the
event yields in the mass range $100\:  < M_\text{col} < 150\GeV$ are shown in Table~\ref{tab:EventYieldTable_100_to_150}.
The different channels and categories are combined  to set a $95\%$ CL  upper limit on the branching
fraction of LFV \PH decay in the  $\Pgm\Pgt$ channel, $\mathcal{B}(\PH\to\Pgm\Pgt)$.

\begin{table*}[hbtp]
 \centering
  \topcaption{Event yields in the signal region,  $100\: <  M_\text{col} < 150\GeV $ after fitting for signal and background. The expected contributions are normalized to an integrated luminosity
of 19.7\fbinv. The LFV Higgs boson signal is the expected yield for $B(\PH \to \mu \tau)=0.84\%$ with the SM Higgs boson cross section.}
  \label{tab:EventYieldTable_100_to_150}
   \cmsTable{\textwidth}{
  \begin{tabular}{lccc|ccc} \hline
        \multirow{2}{*}{Sample}                                & \multicolumn{3}{c}{$\PH \to \Pgm \tauh$}                &     \multicolumn{3}{c}{$\PH \to \Pgm \Pgt_{\Pe}$}     \\ \cline{2-7}
                                              &  0-Jet            & 1-Jet            & 2-Jets               &  0-Jet             & 1-Jet            & 2-Jets  \\ \hline
    misidentified leptons                          &  $  1770 \pm 530$      & $   377 \pm 114$      &  $     1.8 \pm   1.0$&  $    42 \pm  17$    &$    16 \pm   7$      & $     1.1 \pm   0.7$  \\
    $ \cPZ \to \Pgt \Pgt$                        &  $   187 \pm   10$     & $    59 \pm   4$      &  $     0.4 \pm   0.2$&  $    65 \pm   3$    &$    39 \pm   2$      & $     1.3 \pm   0.2$   \\
    $ \cPZ\cPZ,\PW\PW$                                  &  $    46 \pm   8$      & $    15 \pm   3$      &  $     0.2 \pm   0.2$&  $    41 \pm   7$    &$    22 \pm   4$      & $     0.7 \pm   0.2$    \\
    $ \PW\gamma$                                &  \NA  & \NA  &  \NA&  $     2 \pm   2$    &$     2 \pm   2$      & \NA    \\
    $ \cPZ \to \Pe\Pe$ or $\mu \mu$                  &  $   110 \pm  23$      & $    20 \pm   7$      &  $     0.1 \pm   0.1$&  $     1.6 \pm   0.7$&$     1.8 \pm   0.8$  & \NA                  \\
    $\ttbar     $                      &  $     2.2 \pm   0.6$  & $    24 \pm   3$      &  $     0.9 \pm   0.5$&  $     4.8 \pm   0.7$&$    30 \pm   3$      & $     1.8 \pm   0.4$   \\
    $\ttbar   $                      &  $     2.2 \pm   1.1$  & $    13 \pm   3$      &  $     0.5 \pm   0.5$&  $     1.9 \pm   0.2$&$     6.8 \pm   0.8$  & $     0.2 \pm   0.1$   \\
    SM \PH background                       &  $     7.1 \pm   1.3$  & $     5.3 \pm   0.8$  &  $     1.6 \pm   0.5$&  $     1.9 \pm   0.3$&$     1.6 \pm   0.2$  & $     0.6 \pm   0.1$    \\
    sum of backgrounds                        &  $  2125 \pm 530$      & $   513 \pm 114$      &  $     5.4 \pm   1.4$&  $   160 \pm  19$    &$   118 \pm   9$     & $     5.6 \pm   0.9$    \\   \hline
    LFV Higgs boson signal                          &  $    66 \pm  18$      & $    30 \pm   8$      &  $     2.9 \pm   1.1$&  $    23 \pm   6$    &$    13 \pm   3$      & $     1.2 \pm   0.3$    \\   \hline
    data                                      &  $  2147 $             & $   511 $             &  $    10 $           &  $   180 $           &$   128 $             & $     6 $    \\   \hline
  \end{tabular}
  }

\end{table*}

\begin{figure*}[hbtp]\centering
\includegraphics[width=0.48\textwidth]{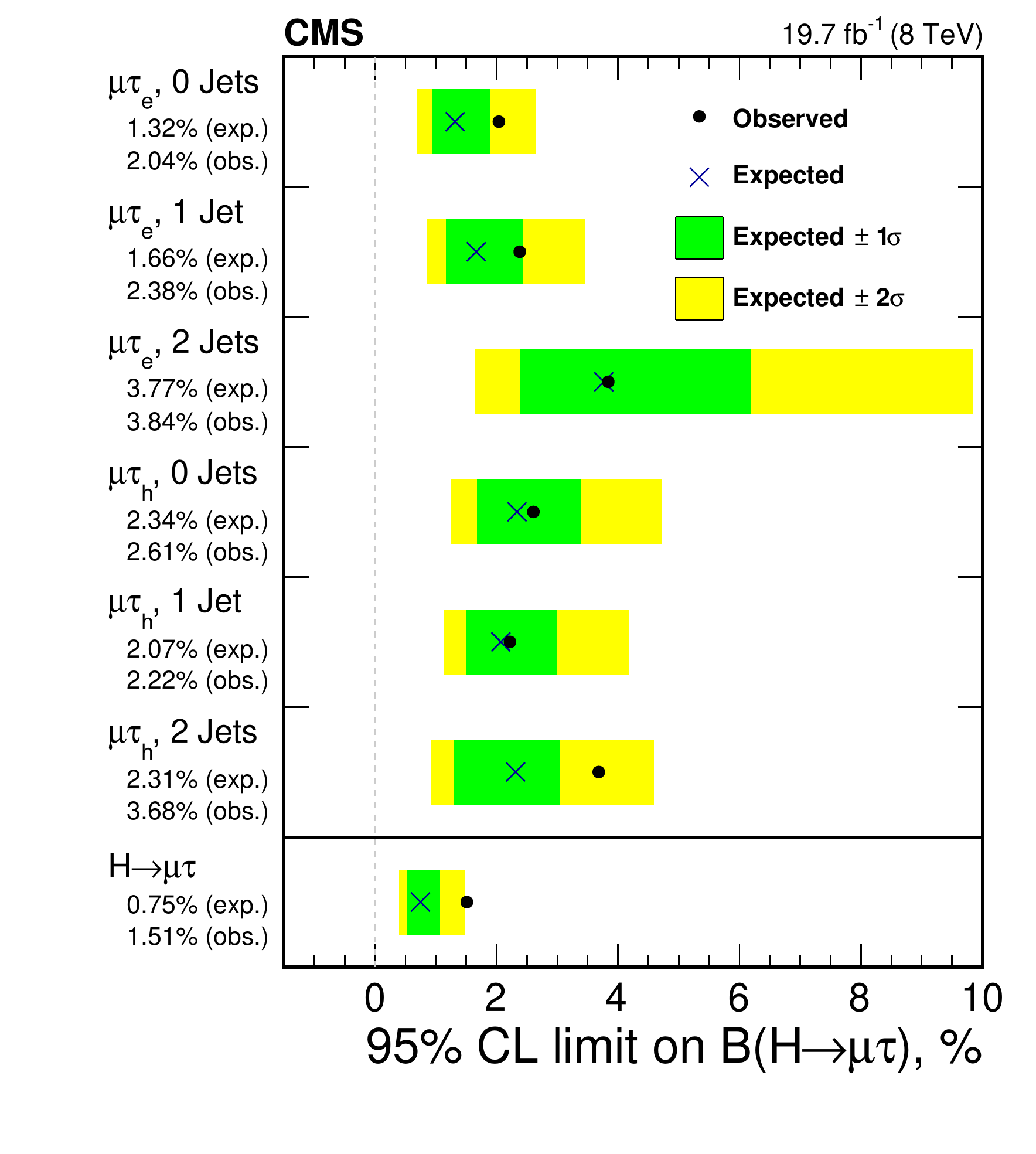}
\includegraphics[width=0.48\textwidth]{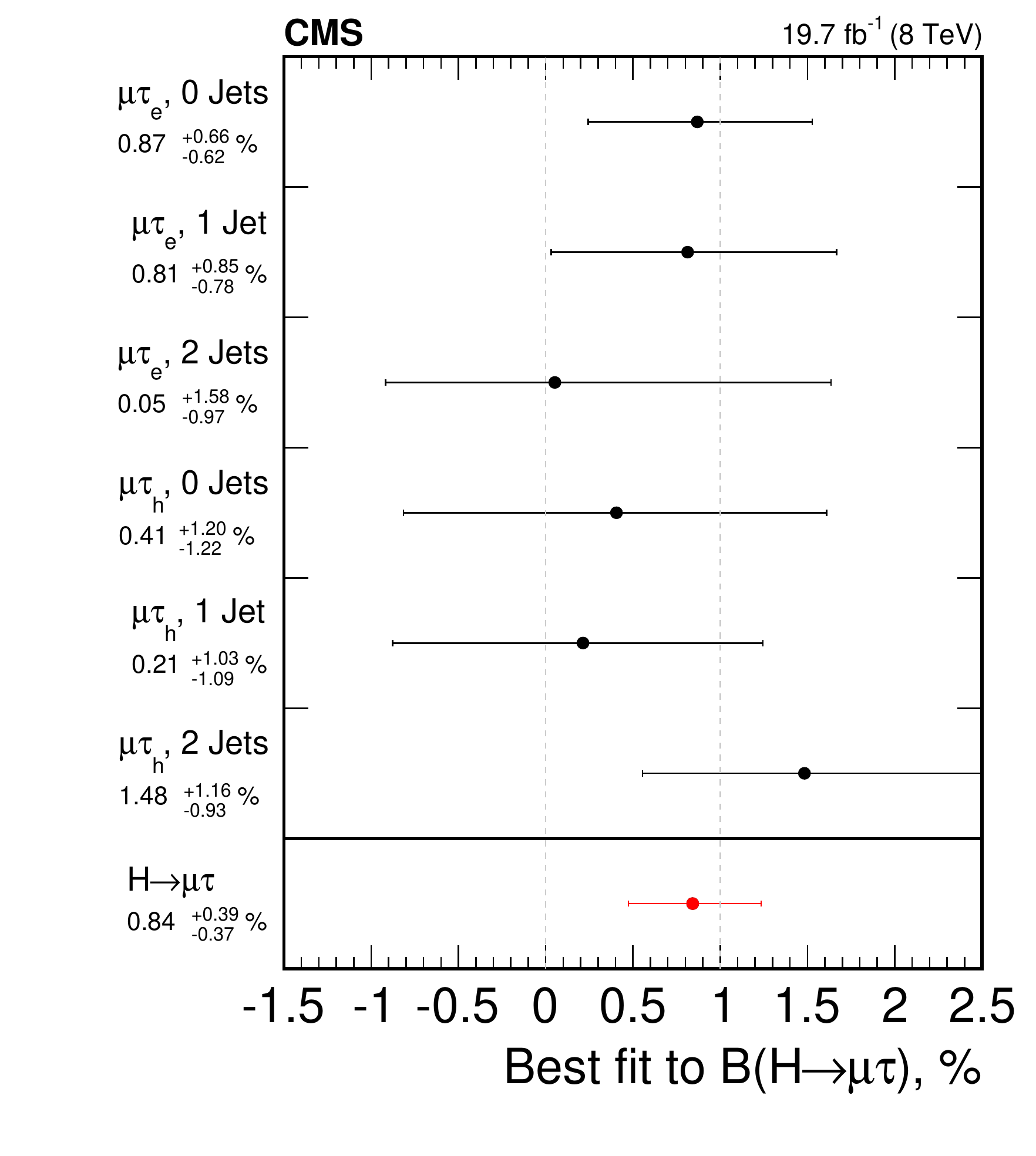}
 \caption{Left: 95\% CL Upper limits by category for the LFV $\PH \to \Pgm \Pgt$  decays. Right: best fit branching fractions by category.}
 \label{fig:limits_summary}\end{figure*}

\begin{figure*}[hbtp]\centering
 \includegraphics[width=0.48\textwidth]{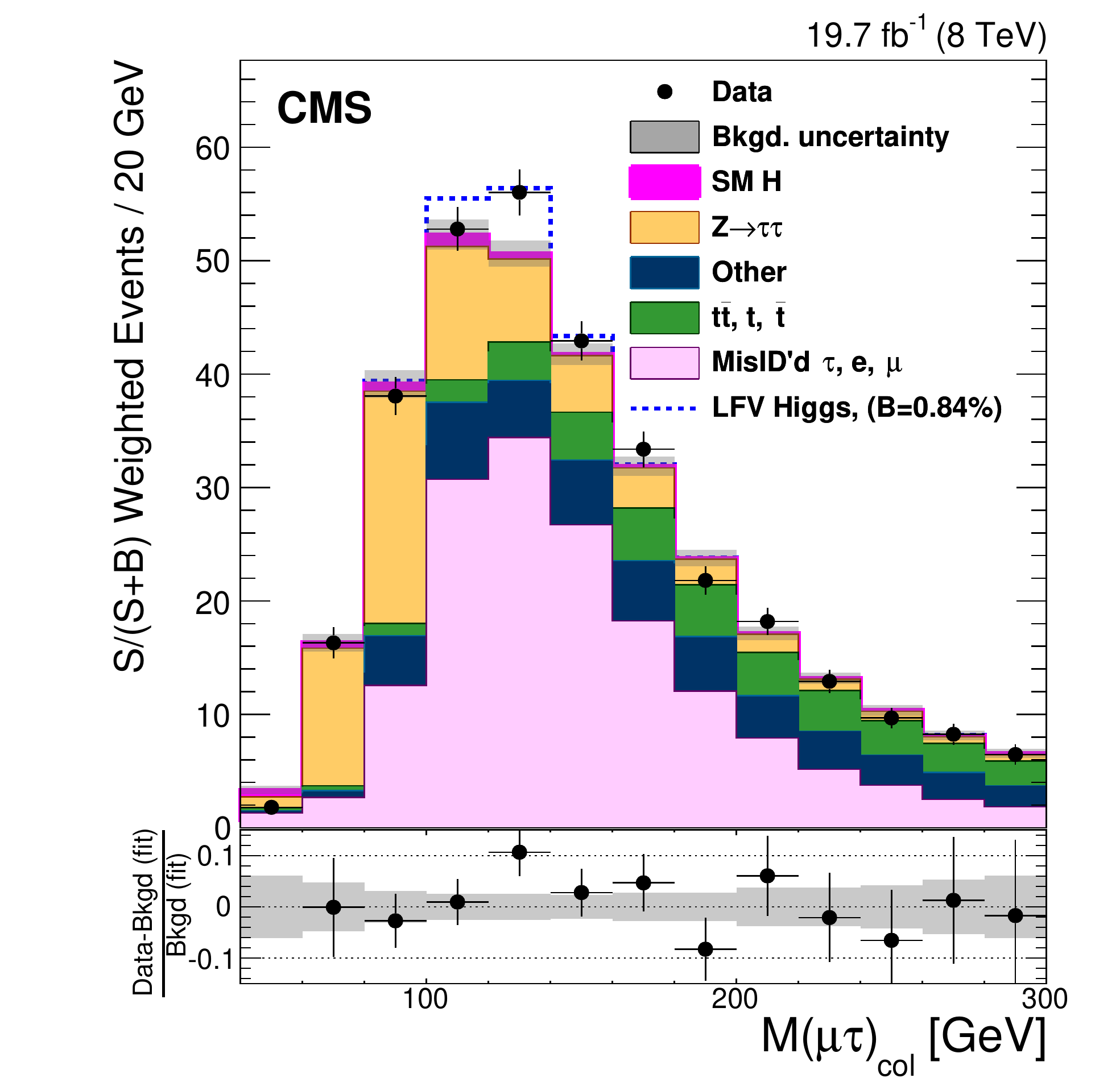}
 \includegraphics[width=0.48\textwidth]{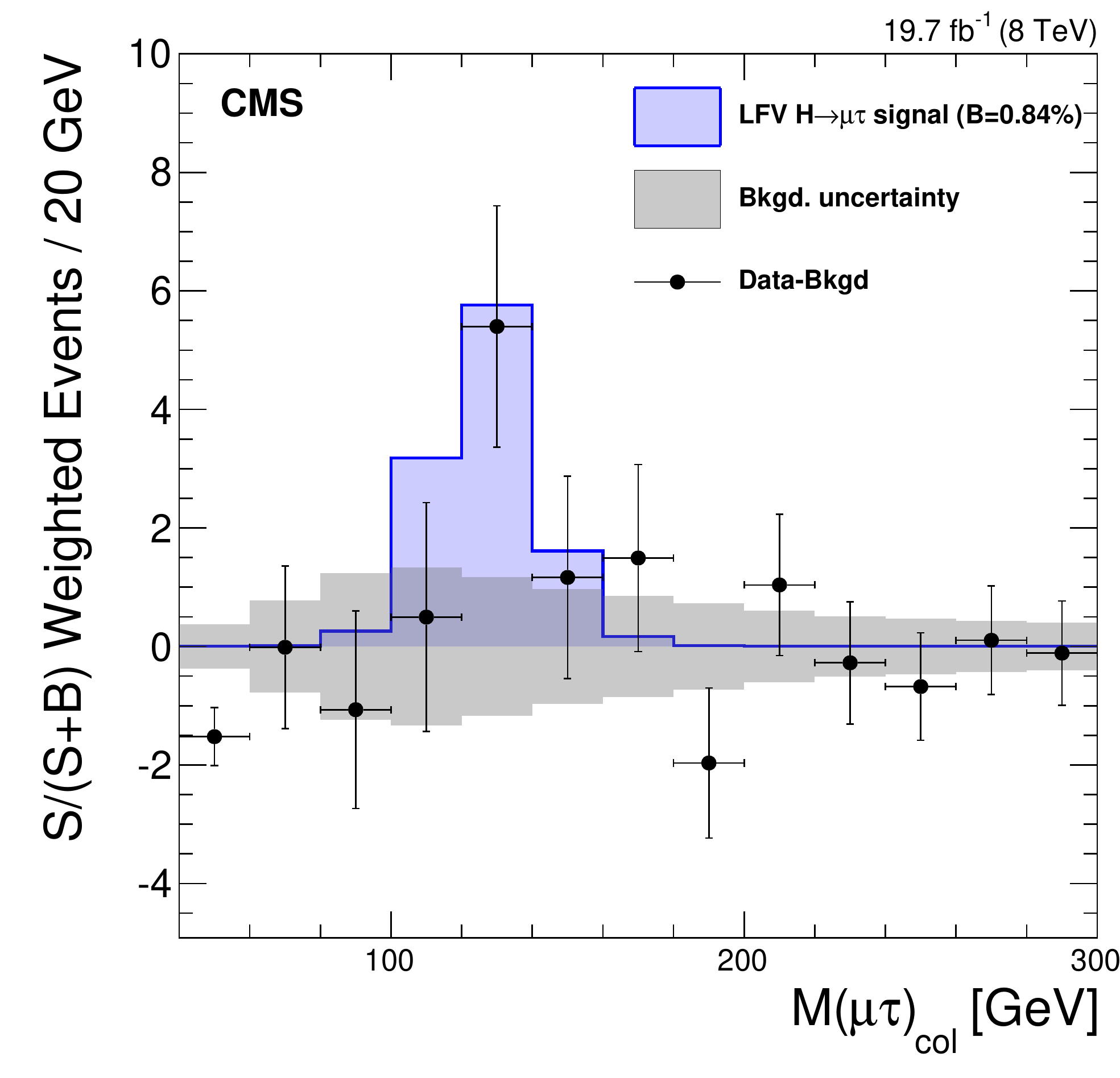}
 \caption{Left: Distribution of $M_\text{col}$ for all categories combined, with each category weighted by significance ($\mathrm{S}/(\mathrm{S}+\mathrm{B})$).
The significance is computed for the integral of the bins in the range $100 < M_\text{col} < 150\GeV$
using $\mathcal{B}(\PH \to \Pgm \Pgt)=0.84\%$. The simulated  Higgs signal shown is for $\mathcal{B}(\PH \to \Pgm \Pgt)=0.84\%$. The bottom panel shows the fractional difference between the observed data and the fitted background.
Right: background subtracted $M_\text{col}$ distribution for all categories combined.}
\label{fig:mcol_all_global_weighted}\end{figure*}

 The observed and the median expected $95\%$ CL upper limits on the $\mathcal{B}(\PH \to \Pgm \Pgt )$ for the \PH mass at 125\GeV are given for each category
 in Table~\ref{tab:expected_limits}.  Combining all
the channels, an expected upper limit of $\mathcal{B}(\PH \to \Pgm \Pgt )<(0.75 \pm 0.38)\%$ is obtained. The
observed upper limit is $\mathcal{B}(\PH \to \Pgm \Pgt ) < 1.51\%$ which is above the expected limit due to an excess of the
observed number of events above the background prediction.
The fit can then be used to estimate the branching fraction if this excess were to be interpreted as a signal.
The best fit values for the branching fractions are given in Table~\ref{tab:expected_limits}.
The limits and best fit branching fractions are also  summarized graphically  in
Fig.~\ref{fig:limits_summary}. The combined categories give a best fit of $\mathcal{B}(\PH \to \Pgm \Pgt )=(0.84^{+0.39}_{-0.37})\%$. The combined excess is 2.4 standard deviations which corresponds to a  $p$-value of 0.010 at $\MH=125$\GeV.
The observed and expected $M_\text{col}$ distributions combined for all channels and categories are shown in
Fig.~\ref{fig:mcol_all_global_weighted}. The distributions are weighted in each channel and category by the $\mathrm{S/(S + B)}$ ratio,
where S and B are respectively the signal and background yields corresponding to the result of the global fit.
The values for S and B are obtained in the  $100 < M_\text{col} < 150\GeV$  region.

\begin{table}[hbtp]
 \centering
  \topcaption{The expected upper limits, observed upper limits and best fit values for the branching fractions for different
    jet categories for the $\PH \to \Pgm \Pgt$  process.
    The one standard-deviation probability intervals around the expected limits are shown in parentheses.}
  \label{tab:expected_limits}
  \cmsTable{\columnwidth}{\begin{tabular}{l|c|c|c} \hline
\multicolumn{4}{c}{Expected Limits} \\ \hline
                       &  \multicolumn{1}{c|}{0-Jet}   & \multicolumn{1}{c}{1-Jet}    &  \multicolumn{1}{|c}{2-Jets}                 \\
                       & (\%)                     & (\%)                     & (\%)                    \\ \cline{2-4}
          $\Pgm\Pgt_{\Pe}$  &  $<$1.32 ($\pm$0.67)   &  $<$1.66 ($\pm$0.85)   &  $<$3.77 ($\pm$1.92)  \\
      $\Pgm\tauh$    &  $<$2.34 ($\pm$1.19)   &  $<$2.07 ($\pm$1.06)   &  $<$2.31 ($\pm$1.18)  \\ \hline
            $\Pgm\Pgt$  &        \multicolumn{3}{c}{  $<$0.75 ($\pm$0.38 ) }                              \\ \hline
\multicolumn{4}{c}{Observed Limits} \\ \hline
          $\Pgm\Pgt_{\Pe}$  &  $<$2.04                &  $<$2.38                &  $<$3.84   \\
      $\Pgm\tauh$    &  $<$2.61                &  $<$2.22                &  $<$3.68   \\ \hline
            $\Pgm\Pgt$  & \multicolumn{3}{c}{  $<$1.51 }   \\ \hline
\multicolumn{4}{c}{Best Fit Branching Fractions} \\ \hline
      \rule[-5pt]{0pt}{17pt}
      $\Pgm\Pgt_{\Pe}$  &  $0.87^{+0.66}_{-0.62}$  &  $0.81^{+0.85}_{-0.78}$  &  $0.05^{+1.58}_{-0.97}$  \\
      \rule[-5pt]{0pt}{17pt}
      $\Pgm\tauh$    &  $0.41^{+1.20}_{-1.22}$  &  $0.21^{+1.03}_{-1.09}$  &  $1.48^{+1.16}_{-0.93}$  \\ \hline
      \rule[-5pt]{0pt}{17pt}
      $\Pgm\Pgt$  & \multicolumn{3}{c}{ $0.84^{+0.39}_{-0.37}$ }   \\ \hline
  \end{tabular}
}

\end{table}

\section{Limits on lepton-flavour-violating couplings}
\begin{figure}[hbt]\centering
\includegraphics[width=0.49\textwidth]{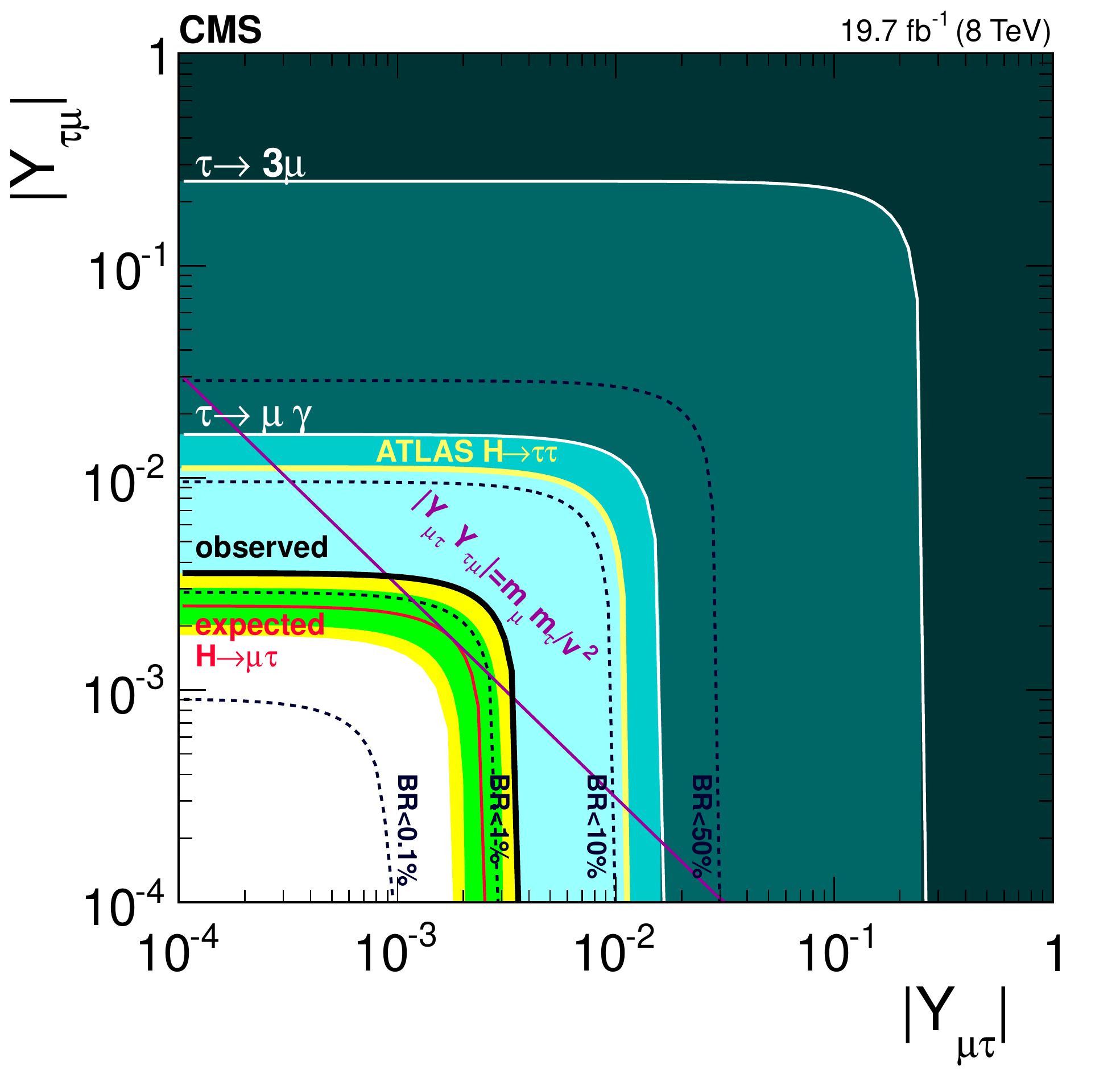}
 \caption{Constraints on the flavour-violating Yukawa couplings, $\abs{Y_{\Pgm\Pgt}}$ and $\abs{Y_{\Pgt\Pgm}}$.
The black dashed lines are contours of $\mathcal{B}(\PH \to \Pgm \Pgt )$ for reference.
The expected limit (red solid line) with one sigma (green)  and two sigma (yellow) bands, and observed limit (black solid line) are derived from the limit on $\mathcal{B}(\PH \to \Pgm \Pgt )$ from the present analysis.  The shaded regions are derived constraints from null searches for $\Pgt \to 3\Pgm$ (dark green) and $\Pgt \to \Pgm \gamma$ (lighter green).
The yellow line is the limit from a theoretical reinterpretation of an ATLAS $\PH \to \Pgt \Pgt$ search~\cite{Harnik:2012pb}.
The light blue region indicates the additional parameter space excluded by our result.
The purple diagonal line is the theoretical naturalness
limit $Y_{ij}Y_{ji} \leq m_im_j/v^2$. }
 \label{fig:yukawalimits}\end{figure}

The constraint on $\mathcal{B}(\PH \to \Pgm \Pgt )$ can be interpreted in terms of LFV  Yukawa couplings~\cite{Harnik:2012pb}.
The LFV decays $\PH \to \Pe\Pgm$, $\Pe\Pgt$, $\Pgm\Pgt$ arise at tree level from the assumed
flavour-violating Yukawa interactions, $Y_{\ell^{\alpha}\ell^{\beta}}$ where $\ell^{\alpha},\ell^{\beta}$ denote the leptons, $\ell^{\alpha},\ell^{\beta}=\Pe,\Pgm,\Pgt$ and $\ell^{\alpha}\neq \ell^{\beta}$.
The decay width $\Gamma(\PH \to \ell^{\alpha}\ell^{\beta})$  in terms of the Yukawa couplings is given by:
\begin{equation*}
\Gamma(\PH \to \ell^{\alpha}\ell^{\beta})=\frac{m_{\PH}}{8\pi}\bigl(\abs{Y_{\ell^{\beta}\ell^{\alpha}}}^2 + \abs{Y_{\ell^{\alpha}\ell^{\beta}}}^2\bigr),
\end{equation*}
and the branching fraction by:
\begin{equation*}
B(\PH \to \ell^{\alpha}\ell^{\beta})=\frac{\Gamma(\PH\to \ell^{\alpha}\ell^{\beta})}{\Gamma(\PH\to \ell^{\alpha}\ell^{\beta}) + \Gamma_{SM}}.
\end{equation*}
The SM \PH decay width is assumed to be $\Gamma_{\mathrm{SM}}=4.1$\MeV~\cite{Denner:2011mq} for $M_{\PH}=125\GeV$.
The 95\% CL constraint on the Yukawa couplings derived from $\mathcal{B}(\PH \to \Pgm \Pgt )<1.51\%$ and the expression for the branching fraction above is:\begin{equation*}
\sqrt{\abs{Y_{\Pgm\Pgt}}^{2}+\abs{Y_{\Pgt\Pgm}}^{2}}<3.6\times 10^{-3}.
\end{equation*}
Figure~\ref{fig:yukawalimits} compares this result to the constraints from previous indirect
measurements.

\section{Summary}
The first direct search for lepton-flavour-violating decays of a Higgs boson to a $\Pgm$-$\Pgt$ pair, based on the full 8\TeV data set collected by CMS in 2012
is presented.  It improves upon previously published indirect limits~\cite{Blankenburg:2012ex,Harnik:2012pb} by an order of magnitude. A slight excess of events with
a significance of $2.4\,\sigma$  is observed, corresponding to a $p$-value of 0.010. The best fit branching fraction is $\mathcal{B}(\PH \to \Pgm \Pgt )=(0.84^{+0.39}_{-0.37})\%$. A
constraint of $\mathcal{B}(\PH \to \Pgm \Pgt )<1.51\%$ at 95\% confidence level is set.  The limit is used to constrain the Yukawa couplings,
$\sqrt{\smash[b]{\abs{Y_{\Pgm\Pgt}}^{2}+\abs{Y_{\Pgt\Pgm}}^{2}}}<3.6\times 10^{-3}$. It improves the current bound by an order of magnitude.

\begin{acknowledgments}
We congratulate our colleagues in the CERN accelerator departments for the excellent performance of the LHC and thank the technical and administrative staffs at CERN and at other CMS institutes for their contributions to the success of the CMS effort. In addition, we gratefully acknowledge the computing centres and personnel of the Worldwide LHC Computing Grid for delivering so effectively the computing infrastructure essential to our analyses. Finally, we acknowledge the enduring support for the construction and operation of the LHC and the CMS detector provided by the following funding agencies: BMWFW and FWF (Austria); FNRS and FWO (Belgium); CNPq, CAPES, FAPERJ, and FAPESP (Brazil); MES (Bulgaria); CERN; CAS, MoST, and NSFC (China); COLCIENCIAS (Colombia); MSES and CSF (Croatia); RPF (Cyprus); MoER, ERC IUT and ERDF (Estonia); Academy of Finland, MEC, and HIP (Finland); CEA and CNRS/IN2P3 (France); BMBF, DFG, and HGF (Germany); GSRT (Greece); OTKA and NIH (Hungary); DAE and DST (India); IPM (Iran); SFI (Ireland); INFN (Italy); MSIP and NRF (Republic of Korea); LAS (Lithuania); MOE and UM (Malaysia); CINVESTAV, CONACYT, SEP, and UASLP-FAI (Mexico); MBIE (New Zealand); PAEC (Pakistan); MSHE and NSC (Poland); FCT (Portugal); JINR (Dubna); MON, RosAtom, RAS and RFBR (Russia); MESTD (Serbia); SEIDI and CPAN (Spain); Swiss Funding Agencies (Switzerland); MST (Taipei); ThEPCenter, IPST, STAR and NSTDA (Thailand); TUBITAK and TAEK (Turkey); NASU and SFFR (Ukraine); STFC (United Kingdom); DOE and NSF (USA).

Individuals have received support from the Marie-Curie programme and the European Research Council and EPLANET (European Union); the Leventis Foundation; the A. P. Sloan Foundation; the Alexander von Humboldt Foundation; the Belgian Federal Science Policy Office; the Fonds pour la Formation \`a la Recherche dans l'Industrie et dans l'Agriculture (FRIA-Belgium); the Agentschap voor Innovatie door Wetenschap en Technologie (IWT-Belgium); the Ministry of Education, Youth and Sports (MEYS) of the Czech Republic; the Council of Science and Industrial Research, India; the HOMING PLUS programme of Foundation for Polish Science, cofinanced from European Union, Regional Development Fund; the Compagnia di San Paolo (Torino); the Consorzio per la Fisica (Trieste); MIUR project 20108T4XTM (Italy); the Thalis and Aristeia programmes cofinanced by EU-ESF and the Greek NSRF; and the National Priorities Research Program by Qatar National Research Fund.
\end{acknowledgments}

\bibliography{auto_generated}

\cleardoublepage \appendix\section{The CMS Collaboration \label{app:collab}}\begin{sloppypar}\hyphenpenalty=5000\widowpenalty=500\clubpenalty=5000\textbf{Yerevan Physics Institute,  Yerevan,  Armenia}\\*[0pt]
V.~Khachatryan, A.M.~Sirunyan, A.~Tumasyan
\vskip\cmsinstskip
\textbf{Institut f\"{u}r Hochenergiephysik der OeAW,  Wien,  Austria}\\*[0pt]
W.~Adam, T.~Bergauer, M.~Dragicevic, J.~Er\"{o}, M.~Friedl, R.~Fr\"{u}hwirth\cmsAuthorMark{1}, V.M.~Ghete, C.~Hartl, N.~H\"{o}rmann, J.~Hrubec, M.~Jeitler\cmsAuthorMark{1}, W.~Kiesenhofer, V.~Kn\"{u}nz, M.~Krammer\cmsAuthorMark{1}, I.~Kr\"{a}tschmer, D.~Liko, I.~Mikulec, D.~Rabady\cmsAuthorMark{2}, B.~Rahbaran, H.~Rohringer, R.~Sch\"{o}fbeck, J.~Strauss, W.~Treberer-Treberspurg, W.~Waltenberger, C.-E.~Wulz\cmsAuthorMark{1}
% ...
\vskip\cmsinstskip
% ...
1:~~Also at Vienna University of Technology, Vienna, Austria\\
2:~~Also at CERN, European Organization for Nuclear Research, Geneva, Switzerland\\
% ...

\end{sloppypar}
\end{document}